\documentclass[useAMS,usenatbib,usegraphicx]{mn2e}

\usepackage{times}
\usepackage{amsmath}
\usepackage{amssymb}
\usepackage{rotating}

\input{psfig.sty}

\newcommand{\AIPS}{{$\cal AIPS\/$}}

\def\gs{\mathrel{\raise0.35ex\hbox{$\scriptstyle >$}\kern-0.6em
\lower0.40ex\hbox{{$\scriptstyle \sim$}}}}
\def\ls{\mathrel{\raise0.35ex\hbox{$\scriptstyle <$}\kern-0.6em
\lower0.40ex\hbox{{$\scriptstyle \sim$}}}}
\def\m@th{\mathsurround=0pt }
\def\eqalign#1{\null\,\vcenter{\openup1\jot \m@th
 \ialign{\strut\hfil$\displaystyle{##}$&$\displaystyle{{}##}$\hfil
 \crcr#1\crcr}}\,}

\title[The nature of the sub-mJy radio population]
      {Deep multi-frequency radio imaging in the Lockman Hole using
	the GMRT and VLA: I.\ The nature of the sub-mJy radio population}
           \author[Ibar et al.]
	     {Edo Ibar,$^{\! 1,2}$\thanks{e-mail: ibar@roe.ac.uk}  
	       R.\,J.\ Ivison,$^{\! 1,2}$ 
               A.\,D.\ Biggs,$^{\! 1,3}$ 
	       D.\,V.\ Lal,$^{\! 4}$
	       P.\,N.\ Best$^{2}$ and  D.\,A.\ Green$^{5}$
               \vspace*{1mm}\\
               $^1$ UK Astronomy Technology Centre, Royal Observatory, 
	       Blackford Hill, Edinburgh EH9 3HJ\\
               $^2$ Institute for Astronomy, University of Edinburgh, 
	       Blackford Hill, Edinburgh EH9 3HJ\\
	       $^3$ European Southern Observatory,
               Karl-Schwarzschild-Str. 2, D-85748 Garching, Germany\\
               $^4$ Institute of Astronomy and Astrophysics, 
	       Academia Sinica. P.O.\ Box 23-141, Taipei 10617, Taiwan\\
	       $^5$ Astrophysics Group, Cavendish Laboratory, 19
               J.\,J.\ Thomson Avenue, Cambridge CB3 0HE
       }

\date{\fbox{\sc Draft dated: \today\ }}

\pagerange{000--000}

\begin{document}

\maketitle

\begin{abstract} In the run up to routine observations with the
upcoming generation of radio facilities, the nature of sub-mJy radio
population has been hotly debated. Here, we describe multi-frequency
data designed to probe the emission mechanism that dominates in these
faint radio sources. Our analysis is based on observations of the
Lockman Hole using the Giant Metre-wave Radio Telescope (GMRT) -- the
deepest 610-MHz imaging yet reported -- together with 1.4-GHz imaging
from the Very Large Array (VLA), well matched in resolution and
sensitivity to the GMRT data ($\sigma_{\rm 610MHz}\sim
15\,\mu$Jy\,beam$^{-1}$, $\sigma_{\rm 1.4GHz}\sim
6\,\mu$Jy\,beam$^{-1}$, {\sc fwhm} $\sim5\,$arcsec). The GMRT and VLA
data are cross-matched to obtain the radio spectral indices for the
faint radio emitters. Statistical analyses show no clear evolution for
the median spectral index, $\alpha_{\rm 1.4GHz}^{\rm 610MHz}$ (where
$S_{\nu}\propto \nu^{\alpha}$), as a function of flux
density. $\alpha_{\rm 1.4GHz}^{\rm 610MHz}$ is found to be
approximately $-0.6$ to $-0.7$, based on an almost unbiased
10-$\sigma$ criterion, down to a flux level of $S_{\rm
1.4GHz}\gs\,100\,\mu$Jy. The fraction of inverted spectrum sources
($\alpha_{\rm 1.4GHz}^{\rm 610MHz}>0$) is less than 10 per cent. The
results suggest that the most prevalent emission mechanism in the
sub-mJy regime is optically-thin synchrotron, ruling out a dominant
flat spectrum or ultra-steep spectrum radio population.  The spectral
index distribution has a significant scatter,
$\Delta\alpha\approx0.4-0.5$, which suggests a mixture of different
populations at all flux levels. Spectroscopic classification of radio
sources with X-ray emission has allowed us to estimate that the
fraction of radio-quiet AGN at $30\,\mu{\rm Jy} \ls S_{\rm
1.4GHz}<\,300\,\mu$Jy is roughly $25\pm10$ per cent, suggesting that
star-forming galaxies dominate the sub-mJy regime.
\end{abstract}


\section{Introduction}

In early studies, radio astronomy was limited to bright sources
associated with rare luminous ($L_{\rm 1.4GHz}\approx
10^{25-29}$\,W\,Hz$^{-1}$) radio galaxies and quasars (QSOs).
Galaxies with nuclear activity are usually characterised by powerful
radio lobes, which are evidence of interactions between highly
collimated relativistic flows -- coming from the nuclear activity --
and the interstellar/intergalactic medium. These magnificent
radio-loud structures were classified by \citet{Fanaroff74} depending
on their shape (FR\,{\sc i} and {\sc ii} classes), and optical
identifications showed that these active galactic nuclei (AGN) are
usually hosted by massive elliptical galaxies \citep{Matthews64}. It
was not until the 1980s that radio source counts at the sub-mJy level
revealed a new radio population \citep{Windhorst85, Mitchell85}. The
nature of the faint radio sources which dominate the number counts
below $\sim1\,$mJy is controversial.  Various studies
\citep[e.g.][]{Simpson06, Seymour08, Smolcic08} have identified this
population with star-forming galaxies (starbursts, spirals or
irregulars) and radio-quiet AGN (faint FR\,{\sc i}, Seyfert galaxies).

The fractions of AGN and star-forming galaxies that contribute to the
sub-mJy radio regime is still hotly debated. Many different approaches
have been taken to disentangle these two populations: using
far-infrared (far-IR)/radio flux ratios \citep{Donley05, Ibar08};
tackling their radio brightness temperatures and luminosities
\citep{Wrobel2005, Garrett2005}; resolving their radio morphologies
\citep{Muxlow05, Biggs08}; identifying optical host galaxies via
morphology \citep{Padovani07}, or spectroscopy \citep{Gruppioni99,
Barger07}, or their locus in colour-colour diagrams \citep{Ciliegi05};
via X-ray identifications \citep{Simpson06}; or using their radio
spectral indices \citep{Richards00, Clemens08}. These approaches tend
to yield substantially different results.

In terms of spectral indices, $\alpha$, star-forming galaxies are
usually considered to have a mean spectral index between $-0.8$ and $
-0.7$ (where $S\propto\nu^{\alpha}$), with a relatively small
dispersion, $0.24$ \citep{Condon92}. A sample of $z<0.5$ FR {\sc i} \&
{\sc ii} sources have also been found to have similar spectral indexes
(between 178\,MHz and 750\,MHz), with mean and scatter of
$\alpha=0.74\pm0.19$ and $\alpha=0.79\pm0.14$, respectively
(\citealt{Laing83}). This implies that studies based on the radio
spectral index have large difficulties disentangling star-forming from
steep-spectrum FR-AGN populations. Nevertheless, the radio spectral
index is sensitive to core-dominated radio-quiet AGN
(\citealt{Blundell07}), GHz-peaked sources (GPS;
\citealt{Gopal-Krishna83,Odea98,Snellen00}) and the ultra-steep
spectrum sources (USS; \citealt{Rottgering94, Chambers96, Jarvis01})
usually found at high redshift.

Recent studies, combining 610-MHz and 1.4-GHz data, have found
evidence for flatter spectral indices \citep{Bondi07, Garn08a} and
larger dispersions at sub-mJy radio fluxes \citep[e.g.][]{Mag08},
suggesting that core-dominated radio-quiet AGN are playing a key role
in the sub-mJy radio population.

In this paper, we present two very deep radio images centred on the
Lockman Hole (LH): the deepest to date at 610\,MHz
($\sigma\sim\,$15\,$\mu$Jy\,beam$^{-1}$) obtained using the Giant
Metrewave Radio Telescope (GMRT), and a deep 1.4-GHz image
($\sigma\sim\,$6\,$\mu$Jy\,beam$^{-1}$) obtained using the Very Large
Array (VLA). At these long wavelengths the dominant powering process
is synchrotron radiation.  We merge the two datasets to characterise
the spectral index of the $\mu$Jy radio population as a function of
flux density, thereby probing the physical mechanisms that dominate in
this enigmatic radio population: optically thin (steep spectrum) or
self-absorbed (hard spectrum) synchrotron emission. Our deep,
well-matched observations -- about three times deeper than previous
data -- mitigate the well-known bias towards the detection of steeper
spectra at longer wavelengths, or flatter spectra at shorter
wavelengths. This work provides a parameterisation of the radio
spectral energy distribution (SED) that can be used to estimate more
precise $k$-corrections for the observed radio emitters.

Throughout this paper we assume a Universe with $\Omega_{\rm m}=0.27$,
$\Omega_\Lambda=0.73$ and $H_0=71$\,km\,s$^{-1}$\,Mpc$^{-1}$.

\section{Radio observations}

\label{observations}

\subsection{GMRT}

During six 12-hr sessions in 2006 February and July we obtained data
for three pointings (primary beam {\sc fwhm}\,$\sim43\,$arcmin) in the
Lockman Hole (see Table~\ref{table_pointings}), separated by
11\,arcmin ({\sc lockman-e, lock-3, lhex4}), typically with 28 of the
30 antennas that comprise the GMRT,\footnote{GMRT is run by the
National Centre for Radio Astrophysics of the Tata Institute of
Fundamental Research.}  near Pune, India. The total integration time
in each field, after overheads, was 16\,hr. We recorded 128 channels
($\Delta\nu=$16\,MHz each) centred on 610\,MHz every 16\,s integration
in the lower and upper sidebands (LSB and USB centred at 602 and
618\,MHz, respectively) in each of two polarisations.  Integrations of
40-min duration were interspersed with 5-min scans of the nearby
calibrator, 1035+564 ($S_{\rm 610MHz}\approx\,$2\,Jy), with scans of
3C\,48, 3C\,147 and 3C\,286 ($S_{\rm 610MHz} = 29.4$, $38.3$ and
$21.1$\,Jy, respectively) for flux and bandpass calibration.

Calibration initially followed standard recipes within \AIPS, using
{\sc fitld, indxr} and {\sc setjy}. However, because of concerns that
some baselines were picking up signal from local power lines and that
1035+564 might be too weak to act as a reliable secondary flux
calibration source, a raft of new measures were introduced to avoid
detrimental effects on the resultant images.

For each session, the bright source least affected by radio-frequency
interference (RFI) and with the fewest malfunctioning antennas was
{\sc split} and chosen to be the primary flux density and bandpass
calibrator. After intensive manual flagging of RFI using {\sc spflg}
and {\sc tvflg}, the chosen calibrator was self-calibrated in
phase. Antenna-based bandpass solutions were determined, copied to the
full dataset and used to determine new gain solutions for the primary
calibrator. The gain and bandpass solutions were then applied to the
entire dataset with no time-dependent corrections. The secondary
calibrator was employed to identify problems with the antennas rather
than to track changes in gain, although a more conventional approach
was used to generate images with good positional information for use
in initial phase self-calibration.

Next, calibrated data were processed with the {\sc flgit}
RFI-rejection algorithm. Each 128-channel integration was split into a
series of seven 15-channel pieces, discarding the first 10 and last 13
channels, and points deviating from linear fits to each piece by more
than 5\,$\sigma$ were rejected. Data brighter than 1.5$\times$ the
brightest calibrator were also rejected, leaving around 70 to 95 per
cent of the original data intact, depending on the severity of the
RFI.

\begin{figure*}
\begin{center}
  \includegraphics[scale=0.8]{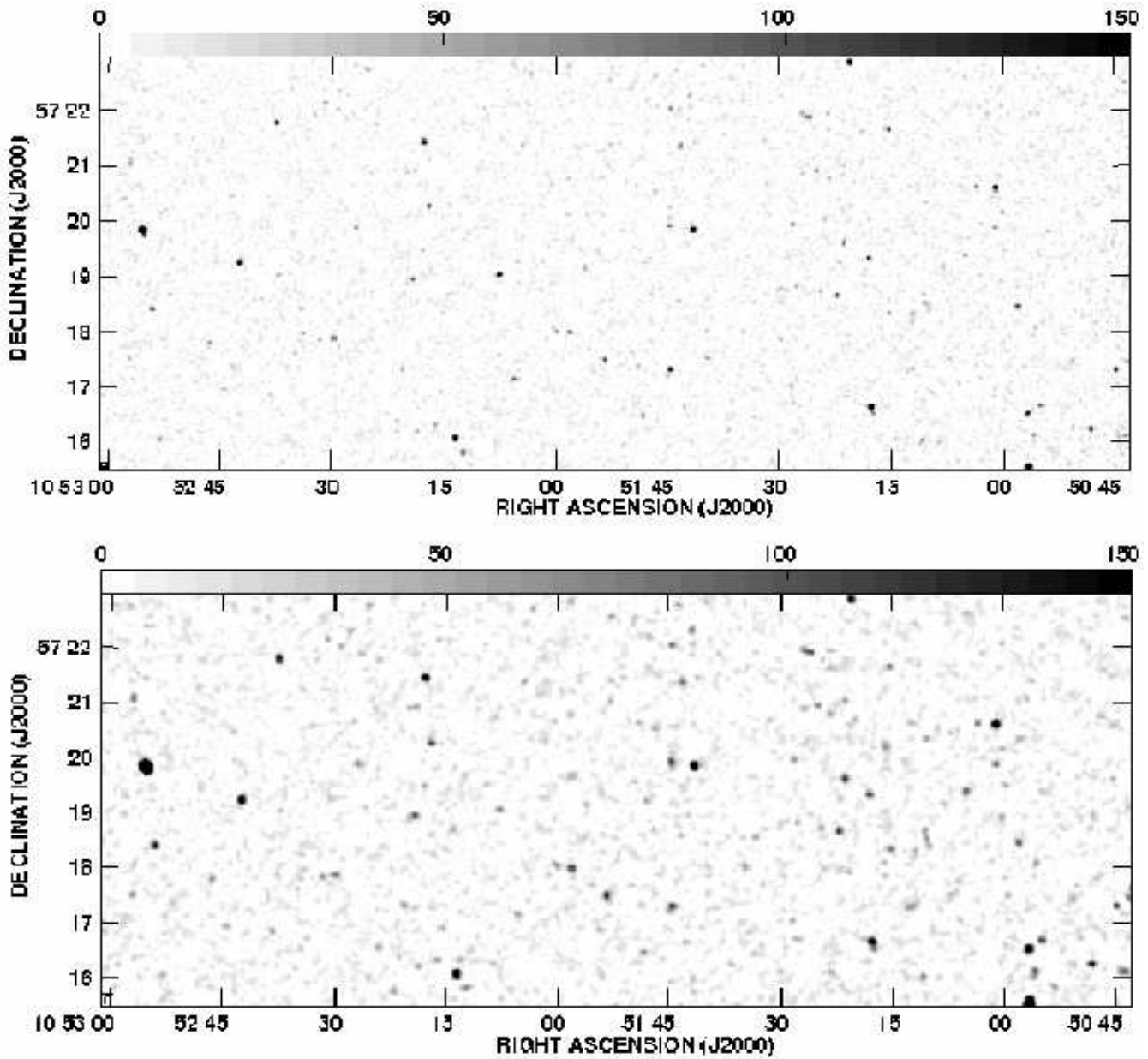}
  \caption{{Top:} A small region (18.7$\times$7.5 arcmin$^2$) near the
  centre of the 1.4-GHz mosaic. The linear grey-scale runs from 0 to
  150\,$\mu$Jy\,beam$^{-1}$. The synthesised beam is
  4.3\,$\times$\,4.2 arcsec$^2$ ({\sc fwhm}) with a position angle of
  77$^{\circ}$. {Bottom:} The deepest 610-MHz image obtained to date,
  covering the same region and with the same linear grey-scale as the
  VLA image on top. The brightness scale of the image has been
  multiplied by $0.56$ ($=[1400/610]^\alpha$, where $\alpha=-0.7$) in
  order to visually compare it with the 1.4-GHz map. The restoring
  beam measures 7.1\,$\times$\,6.5 arcsec$^2$ ({\sc fwhm}) with a
  position angle of 70$^{\circ}$.
  \label{lockmap}
  }
\end{center}
\end{figure*}

The resulting data, now somewhat cleaner, were averaged down to yield
15 channels in each polarisation for each session, pointing and
sideband: a total of 12 dual-polarisation, single-sideband,
single-source datasets.

Before imaging, a specially modified version of {\sc uvavg} (now
standard) was employed to determine and subtract the average value for
each baseline and channel throughout the entire session (hence the need
for time-independent calibration in the preceding steps).

Imaging each of these datasets entailed the creation of a mosaic of 37
facets, each 512$^2$\,pixels (1.25$^2$-arcsec$^2$ per pixel), to cover
the primary beam. A further 6--12 bright sources outside these central
regions, identified in heavily tapered maps, were also imaged. Our aim
was to obtain the best possible model of the sky. {\sc clean} boxes
were placed tightly around all radio sources for use in
self-calibration, first in phase alone ({\sc solmode = `a!p'}), then in
amplitude and phase ({\sc solmode = `a\&p'}), with a solution interval
of 2\,min, staggered by 1\,min. The $uv$ data were weighted using {\sc
robust} = $-$0.5, {\sc uvrange} = 0.8, 100\,k$\lambda$ and {\sc
uvtaper} = 28, 28\,\,k$\lambda$ with {\sc uvbox} = 5.

After {\sc clean} components were subtracted from the $uv$ data, more
manual flagging was applied, as well as another pass through the {\sc
uvavg} task and a clip at the 350-mJy level using {\sc uvflg} (now
{\sc clip}).  {\sc clean} components were re-introduced ({\sc uvsub,
factor=}$-$1), then data with common sidebands from February and July
were combined using {\sc dbcon} to reduce the variation in beam size
and shape amongst the datasets.

The final six mosaics (see Table~\ref{table_pointings}), two for each
pointing (LSB and USB), were then convolved to a common beam size
(7.1\,arcsec $\times$ 6.5\,arcsec, with the major axis at position
angle 70$^{\circ}$) The noise (see Table~\ref{table_pointings}) from
each image is estimated and introduced in their headers using {\sc
imean}, before all are knitted together using {\sc flatn}.  An
appropriate correction was made for the attenuation of the primary
beam, with data rejected at radii beyond where the gain drops to 30
per cent, i.e. at a distance of $\sim28\,$arcmin from the pointing
centre. This final image has a noise level in the central
$\sim\,$100\,arcmin$^2$ of 14.7\,$\mu$Jy\,beam$^{-1}$, the deepest map
so far reported at 610\,MHz, despite the modest integration time
(16\,hr on sky for each pointing).

Based on the brightest pixel in the mosaic, $0.032\,$Jy\,beam$^{-1}$,
we reach a {\it dynamic range} between $\sim\,2,200:1$ and $900:1$
considering the central and local r.m.s.\ noise, respectively. Our
images, however, may be long way from being limited by dynamic
range. \citet{Garn08b} reaches $\sim$9,000:1 and it appears likely
that we could push down to the confusion limit in the observations
discussed in this paper.

\subsection{VLA}

New and archival data were obtained in the same three positions using
the National Radio Astronomy Observatory's (NRAO\footnote{NRAO is
operated by Associated Universities Inc., under a cooperative
agreement with the National Science Foundation.}) VLA, largely in its
B configuration. At 1400\,MHz this yielded images (primary beam {\sc
fwhm}\,$\sim32\,$arcmin) well matched to the resolution of GMRT.  We
tapered our A- and B-configuration data in the {\sc lockman-e} field
\citep{Ivison02} to yield images with a near-circular 4.0-arcsec
synthesised beam. Using the same techniques outlined earlier, we then
combined this central field with images made using pure
B-configuration data in the two other pointings: the designated {\sc
lock-3}, 11\,arcmin to the south west, for which we obtained 17\,hr of
data in 2005 March \citep{Ivison07}; plus archival data for {\sc
lhex4}, 11\,arcmin to the north east of {\sc lockman-e}, which
comprises 31\,hr of integration (see Table~\ref{table_pointings}). The
final mosaic-image has an {\rm r.m.s.} in the central 100\,arcmin$^2$
of $6.0\,\mu$Jy\, beam$^{-1}$, and a resolution of 4.3\,$\times$\,4.2
arcsec$^2$ ({\sc fwhm}) at position angle of 77$^{\circ}$.

\begin{table}
\scriptsize
\begin{center}
\begin{tabular}{|cccc|}
\hline
\multicolumn{4}{c}{GMRT pointings}\\
Field & R.A. (hr:min:sec) & Dec. (deg:min:sec) & 
{\rm r.m.s.} ($\mu$Jy\,beam$^{-1}$) \\
\hline
{\sc lhex-4}  & 10:52:56.0 & +57:29:06.0 & 33.7 \quad (USB) \\
 & & & 29.6 \quad (LSB) \\
{\sc lockman-e} & 10:51:59.0 & +57:21:28.2 & 26.2 \quad (USB) \\
 & & & 26.0 \quad (LSB) \\
{\sc lock-3}  & 10:51:02.0 & +57:13:50.4 & 24.5 \quad (USB) \\
 & & & 23.7 \quad (LSB) \\
\hline
\multicolumn{4}{c}{VLA pointings}\\
Field & R.A. (hr:min:sec) & Dec. (deg:min:sec) & {\rm r.m.s.} ($\mu$Jy\,beam$^{-1}$) \\
\hline
{\sc lhex-4} & 10:52:56.0 & +57:29:06.0 & 7.2 \\
{\sc lockman-e} & 10:52:08.8 & +57:21:33.8 & 7.6 \\
{\sc lock-3} & 10:51:02.0 & +57:13:50.4 & 11.0 \\
\hline
\end{tabular}
\end{center}
\caption{The GMRT and VLA pointings used in this work. USB and LSB
         correspond to the upper and lower side bands,
         respectively.
\label{table_pointings}
}
\end{table}

\section{Catalogues}
\label{cata}

Initially, we extracted sources down to a peak-to-local-noise
ratio\,$=3$ (hereafter PNR), in order to identify all possible faint
and/or extended emission. The sources included in the final catalogues
were selected to have PNR\,$\ge$\,5.

\subsection{Initial source extraction}
\label{sext}

Sources were extracted from the final {\sc flatn}ed mosaics (images of
Stokes $I$), down to a 3-$\sigma$ (thereafter $\sigma$ refers to the
local noise) peak level, using the \AIPS\ routine, {\sc sad} ({\sc
cparm} = 500, 100, 50, 10, 6, 4, 3, 2.5; {\sc dparm}(1)=3; {\sc
dparm}(2)= [15$\,\mu$Jy for GMRT; 7$\,\mu$Jy for VLA]; {\sc icut =
0.1}; {\sc gain = 1}). To provide a reliable noise-based extraction
criterion, a noise map was generated from the Stokes $I$ image using
{\sc rmsd} ({\sc imsize}=71,-1; {\sc optype=`hist'}). This noise map
was introduced as a secondary image for {\sc sad} ({\sc dparm(9) =
3}), which ensured reliable source extraction around bright sources
and near the map edges (see \S\ref{com_garn}). The increasing
uncertainties in the {\sc sad}-Gaussian fits toward faint PNRs may
result in sources having smaller areas than the beamsize (see the
smallest {\sc cparm} parameter). We use a threshold in peak flux
density instead of a threshold in integrated flux because peak flux
density is a linearly independent variable in the {\sc sad} fitting
procedure (actually, in {\sc jmfit}), whereas integrated flux density
is a product of peak flux and source area. This translates into
cleaner and more complete selection criteria. Note, however, that
evidences for an anticorrelation between peak fluxes and source area
have been found by \citet{Condon97} in images with uncorrelated
noise. \citeauthor{Condon97} shows that this effect disappear when the
noise is correlated at similar scales than the synthesised beam. Given
by the convolution to a common beamsize made before using {\sc flatn},
we expect no anticorrelation in the Gaussian fit parameters.

\subsection{Instrumental effects}

Four important instrumental effects must be taken into consideration.

\subsubsection{Bandwidth smearing}
\label{subsub_bs}

\begin{figure}
\begin{center}
\includegraphics[scale=0.34]{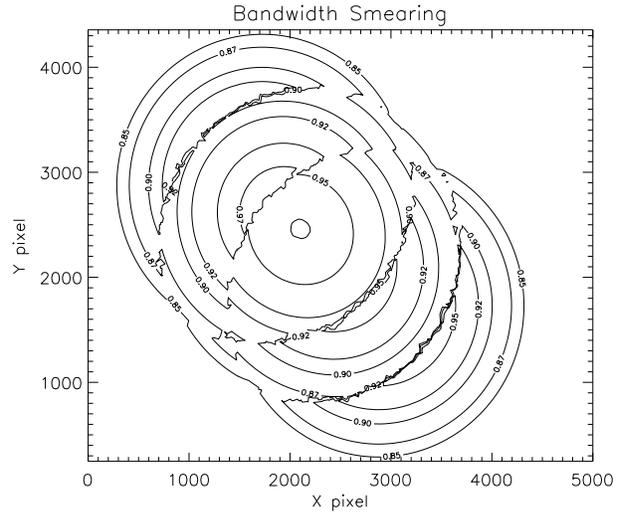}
\caption{Contour plot for the expected point-source bandwidth-smearing
  in the 1.4-GHz mosaic (see Eq.~\ref{mean_bwsc}). Contours are
  plotted at:\, 0.85, 0.87, 0.90, 0.92, 0.95 and 0.97. Note that the
  smallest corrections do not occur exclusively at the image centre.
  \label{sme-bias}
}
\end{center}
\end{figure}

Bandwidth smearing is inevitable when using channels with finite
bandwidth. This affects the peak flux densities of sources, but not
their integrated fluxes. We have estimated this bias through knowledge
of the distance between each source and the different phase centres
and use a theoretical correction given by:

\begin{equation}
  \frac{I}{I_0}=\frac{1}{\sqrt{1+\beta^2}}\hspace{0.5cm}{\rm where}
  \hspace{0.5cm}\beta=\frac{\Delta\nu}{\nu}\frac{\theta}{\theta_{\rm
      syn}}
\end{equation}

\noindent
which is valid for point sources, and assumes a Gaussian bandpass and
circular tapering in the $uv$ plane \citep{Taylor99}. $I$ and $I_0$
are the observed and intrinsic peak brightness, $\Delta\nu$ is the
channel bandwidth, $\nu$ is the frequency of the bandpass centre,
$\theta_{\rm syn}$ is the synthesised beamwidth and $\theta$ is the
distance between the source and the phase centre. This estimation is
in agreement with the {\sc jmfit} correction when {\sc bwsmear} is set
to the channel bandwidth divided by the center frequency.

In overlapping regions we have averaged the correction by using a
minimum variance weighting scheme (see Fig.~\ref{sme-bias}),

\begin{equation}
  \left\langle \frac{I}{I_0}\right\rangle = 
  \frac{\sum_i\frac{I_i/I_0}{\rm
      r.m.s._i^2}} {\sum_i \frac{1}{\rm r.m.s._i^2}},
  \label{mean_bwsc}
\end{equation}

\noindent which takes into account the noise for each pointing before
primary beam correction. This bias was found to be important in the
VLA image and not always negligible for the GMRT data. The maximum
correction factor was $I/I_0\approx0.84$ and $I/I_0\approx0.94$ for
the VLA and GMRT mosaics, respectively. For example,
Fig.~\ref{sme-bias} shows the bandwidth smearing expected in the VLA
mosaic. Note this smearing can be used to correct the peak fluxes in
the final catalogues (see column 5 in Table~\ref{tableGMRT} and
\ref{tableVLA}), but the selection criterion remained unaffected since
it was based on the observed peak values (uncorrected surface
brightness, in Jy\,beam$^{-1}$).

\begin{figure*}
\begin{center}
\includegraphics[scale=0.41]{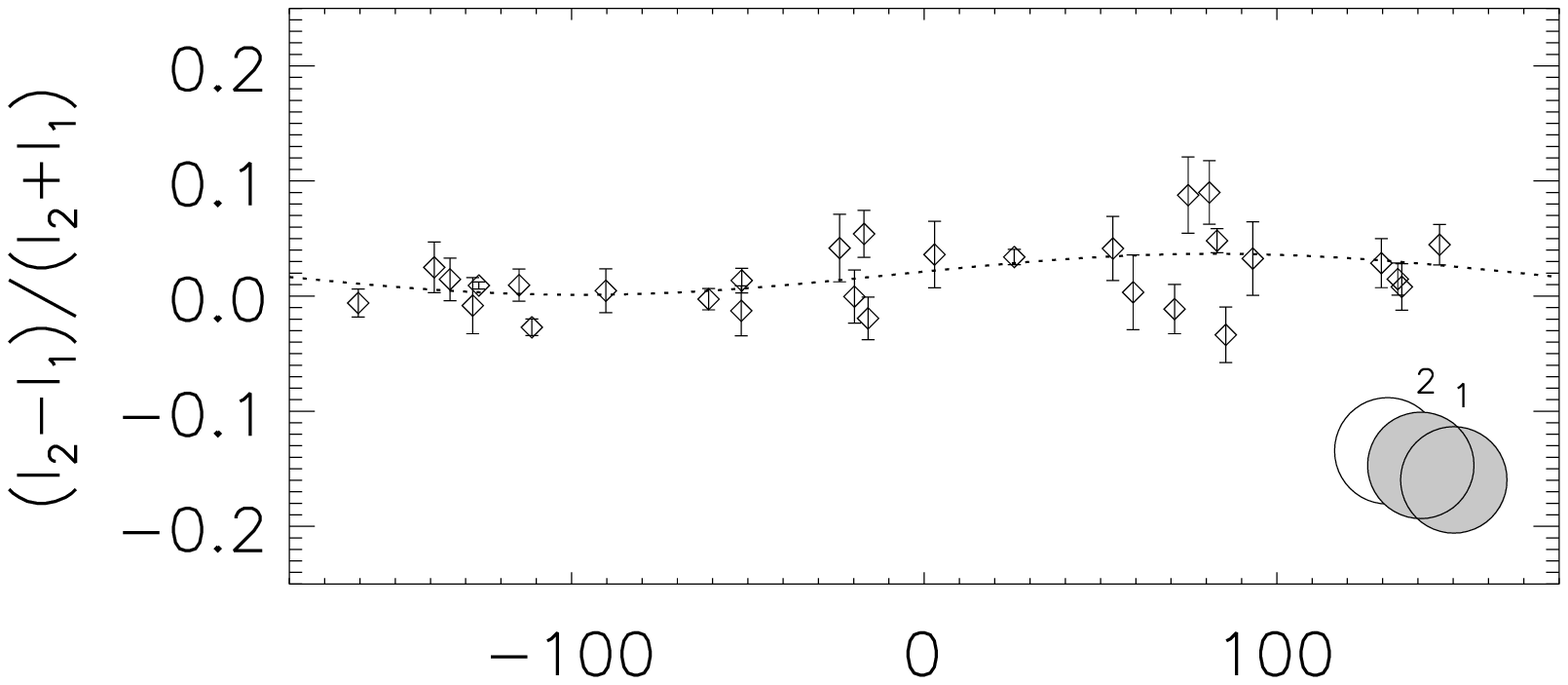}
\includegraphics[scale=0.41]{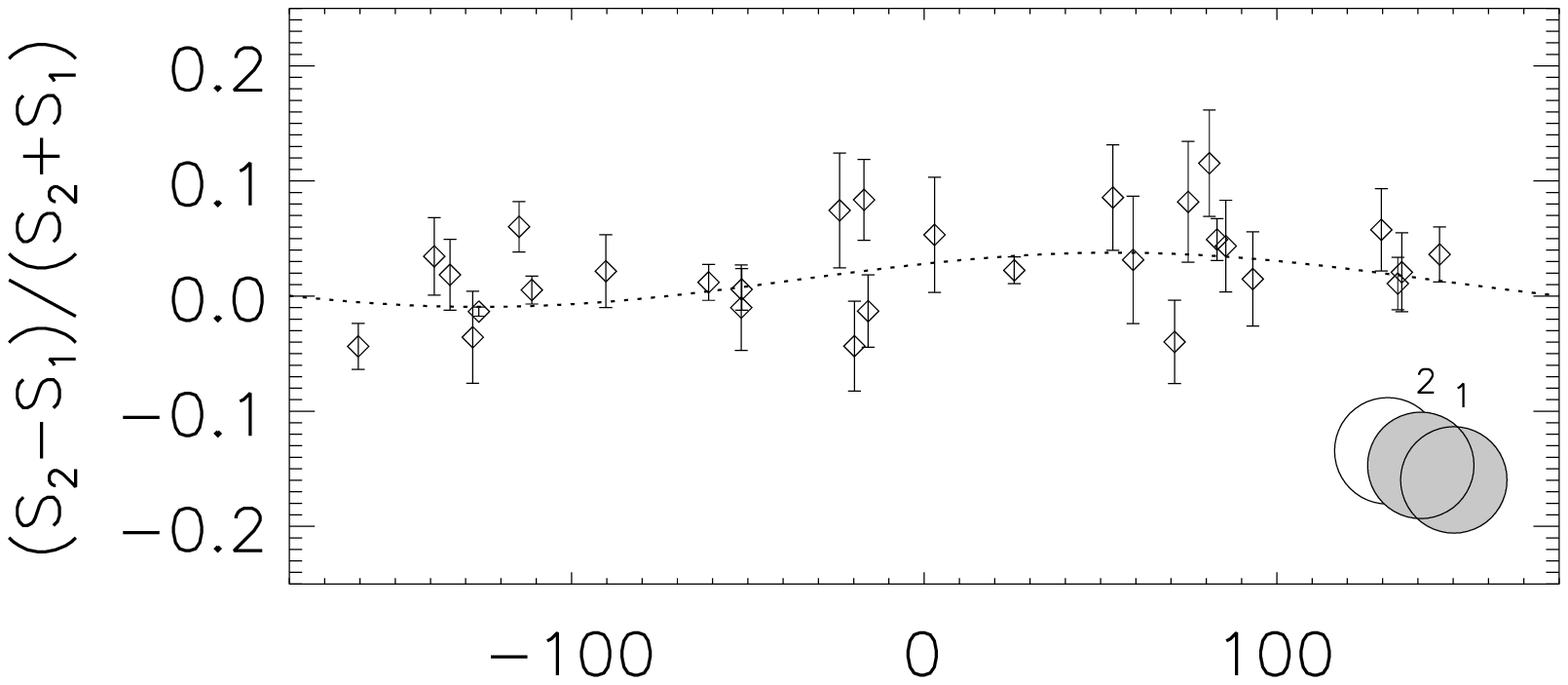}
\includegraphics[scale=0.41]{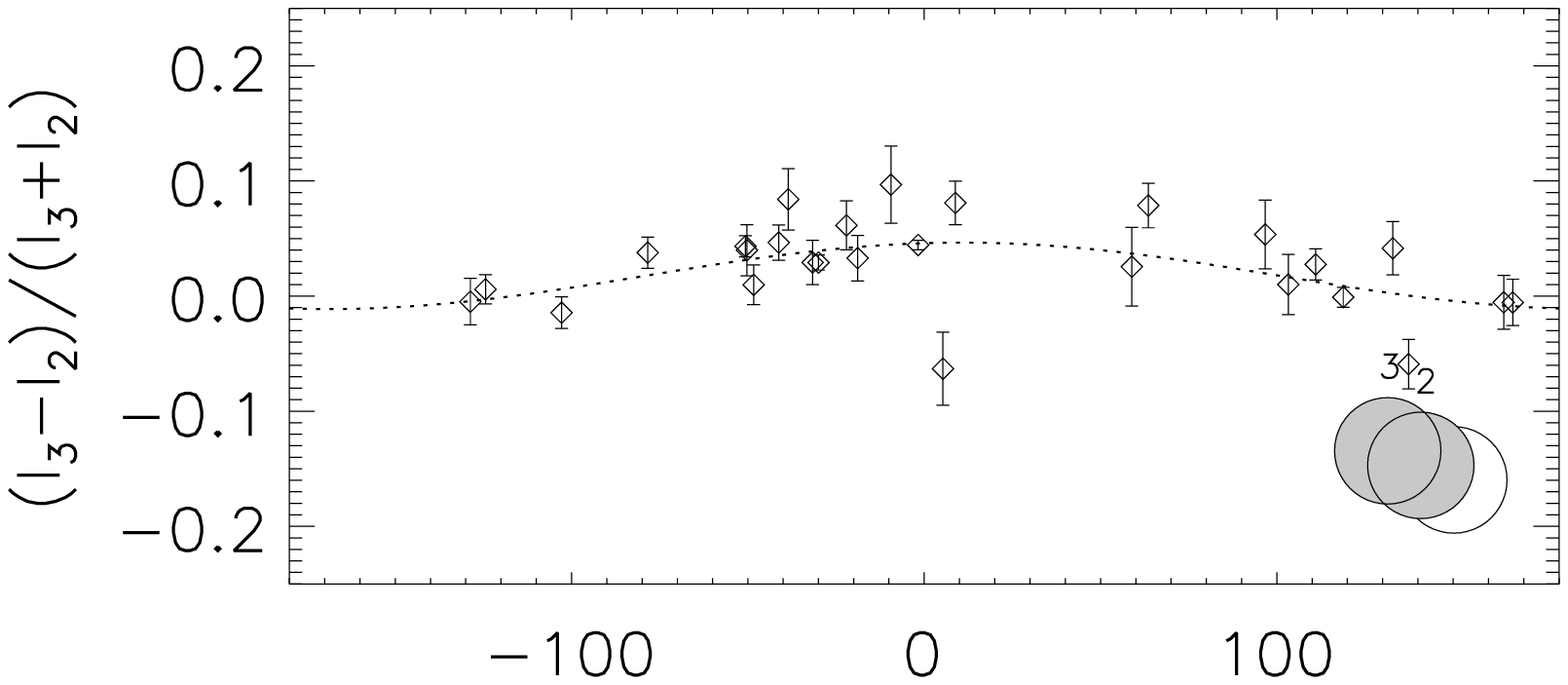}
\includegraphics[scale=0.41]{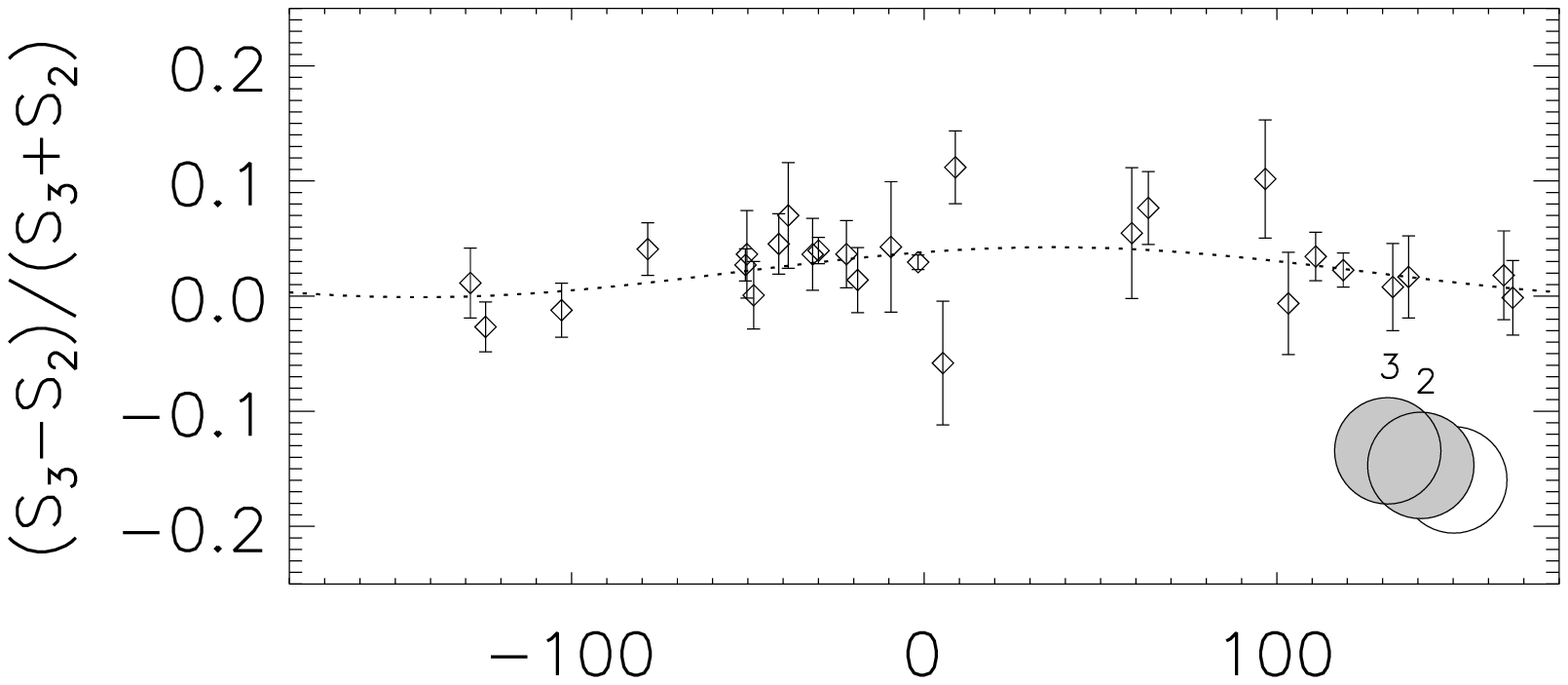}
\includegraphics[scale=0.41]{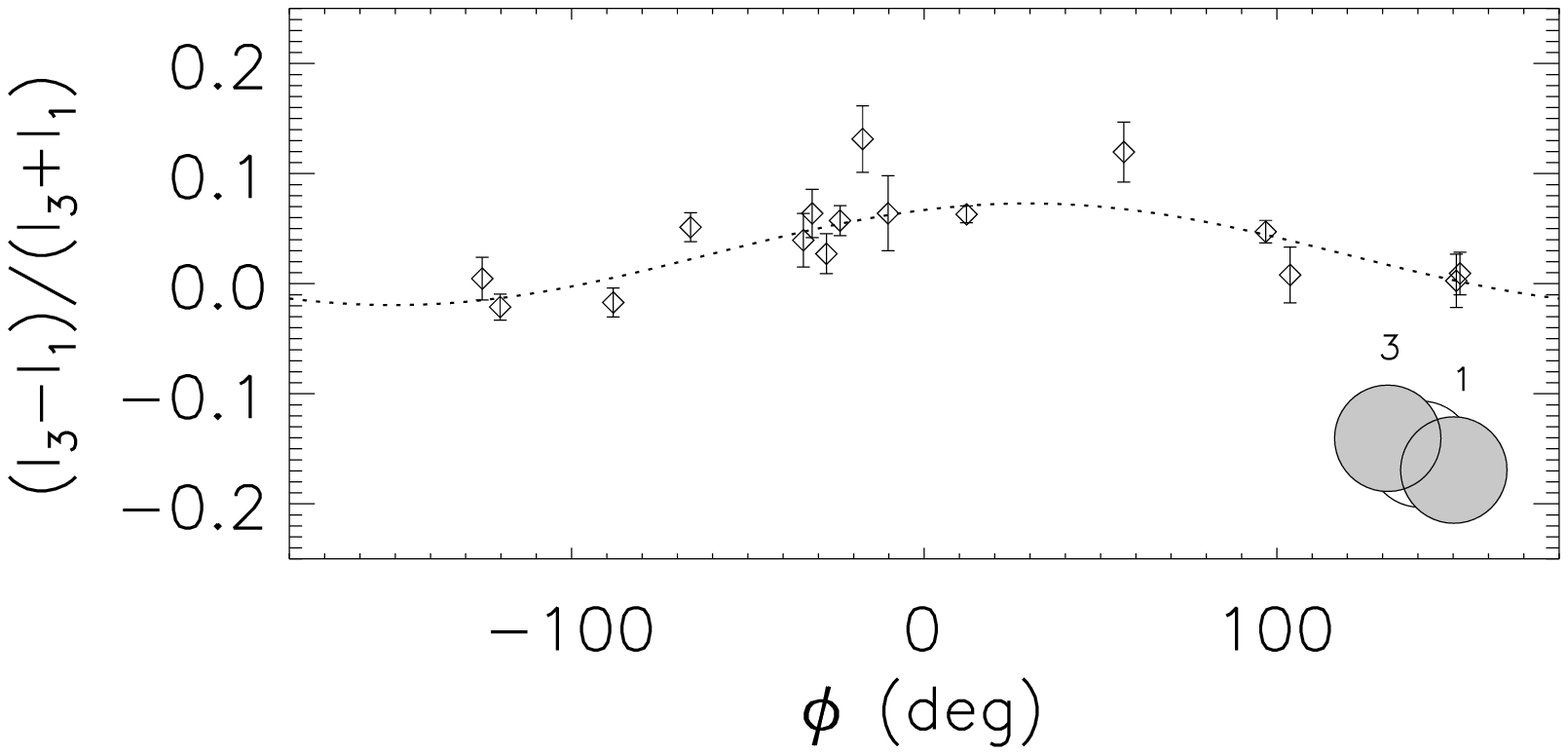}
\includegraphics[scale=0.41]{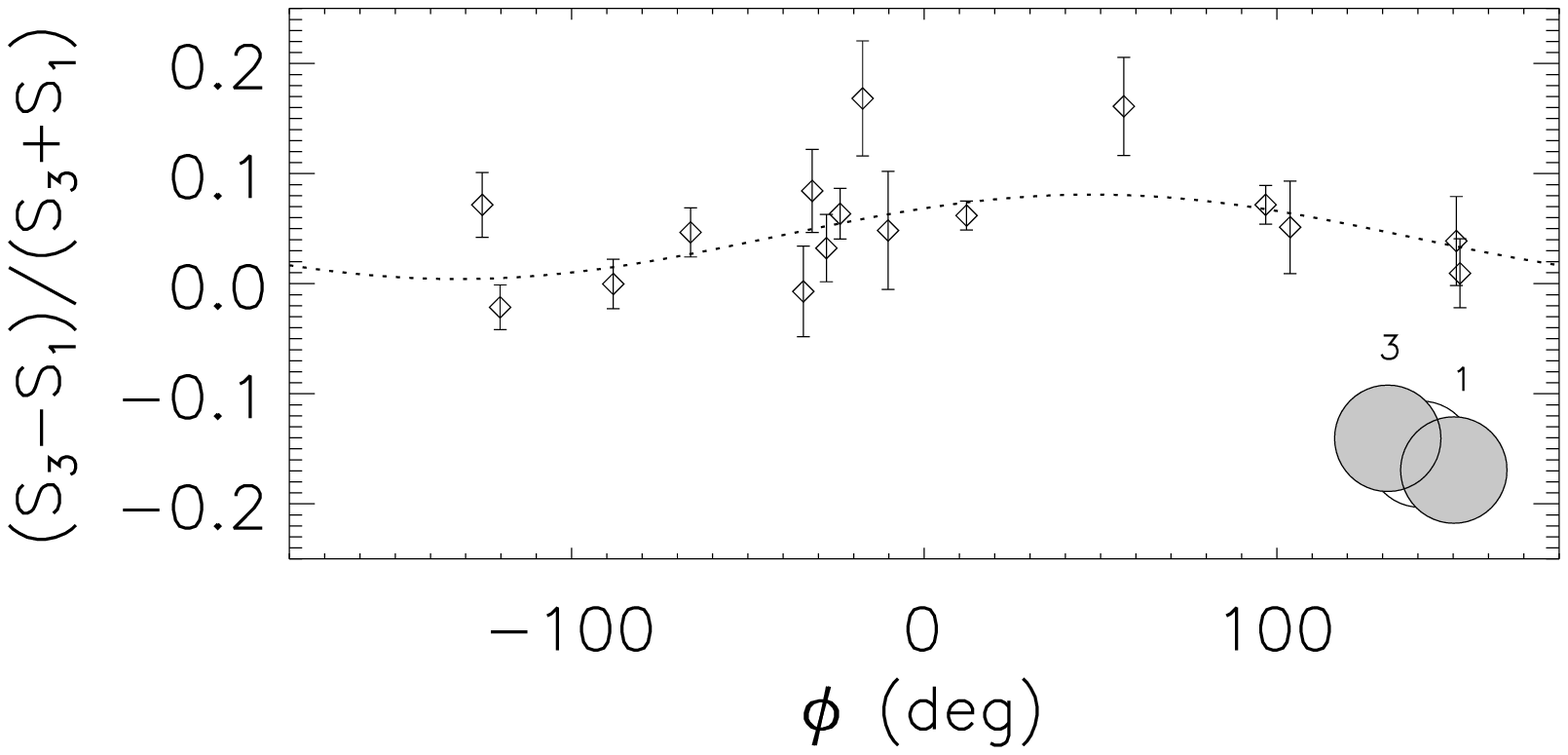}
\caption{These figures compare source estimates from different
  pointings in overlapping regions. The $y$-axis shows the variation
  in peak brightness corrected by bandwidth smearing ($I$ -- {\it
  left}) and integrated flux densities ($S$ -- {\it right}), for all
  sources detected with PNR\,$>25$. Subindexes show the pointings
  being compared (see the bottom-right sketches). The $x$-axis is the
  angle $\phi$ defined in spherical coordinates by the source
  position, the middle distance between the two pointings and the
  northerly direction. Dashed lines are sinusoidal $\chi^2$ fits that
  can provide the direction and amplitude of a primary beam offset
  with respect to the phase centre.
\label{pri-beam1}
}
\end{center}
\end{figure*}

\subsubsection{3-D smearing}

The general response equation for a two-element interferometer is
usually approximated to a 2-D Fourier transform due to the simplicity
of the inversion problem. Nevertheless, this is only valid for
co-planar baselines (where the $w$ axis lies in the direction of the
celestial pole) and for sufficiently small fields of view
\citep{Taylor99}. At low frequencies, the primary beam is large so
this approximation becomes inefficient and a formal 3-D Fourier
transform is required. A smearing effect is expected from the rotation
of the tangential plane with respect to the celestial sphere when the
separation between these two is large.

We can estimate the effect of 3-D smearing in our images. For each VLA
and GMRT pointing image, we have diameters of $\sim\,$45 and
$\sim\,$57\,arcmin, respectively, spanned by seven facets.  The
maximum separation between the tangent plane and the celestial sphere
is given by $\Delta=1-\cos\phi$, where $\phi$ is half the subtended
angle of each facet ($\sim\,$6.4 and $\sim\,$8.1\,arcmin,
respectively). In our images the maximum separation between each
tangent facet and the celestial sphere is $\sim\,$0.09\,arcsec at
1.4\,GHz and $\sim\,$0.14\,arcsec at 610\,MHz. In both cases, this
separation is equivalent to approximately 2 per cent of the
synthesised beamwidth. We consider this bias negligible and no
correction to the observed data was made to correct for 3-D smearing.

\subsubsection{Time-delay smearing}

This smearing is due to the rotation of celestial sources with respect
to the phase tracking centre during the integration time, i.e.\ longer
sampling times lead to more blurred images.

In our observations we used 16- and 5-s integration times to collect
data from the GMRT and VLA, respectively.  Considering theoretical
assumptions (again, see \citealt{Taylor99}), we find that the maximum
loss in peak intensity (expected at the edge of the field of view) is
$\ls2$ per cent for GMRT and $\ls0.3$ per cent for the
VLA. Consequently, we do not implement any correction to the data for
time-delay smearing.

\subsubsection{Primary beam attenuation}

Primary beam attenuation is the intrinsic loss in gain as a function
of distance from the pointing centre due to the antenna response. VLA
images were corrected using the default 10th order fit to the beam
response at 1.4\,GHz, described in \AIPS\ ({\sc explain pbparm}). For
the GMRT images we used the 8th order polynomial fit reported by
N.\,G.\ Kantharia\footnote{\tt
www.ncra.tifr.res.in/$\sim$ngk/primarybeam/beam.html}.

Based on a GMRT mosaic composed of 7 pointings, \citet{Garn07}
reported the primary beam centre was affected by an offset with
respect to the phase centre -- these two are usually coincident. They
revealed a systematic difference between the apparent brightness of
sources observed by adjacent pointings, and solved this problem by
using a common offset of $\sim\,$2.5\,arcmin for the primary
beam corrections.

In order to tackle the offset, first we checked that the images
created by the upper and lower sidebands (IFs -- LSB and USB) from
each of our three pointings are giving consistent results for the
source estimations. We do not find deviations besides of typical
differences in flux calibration of $\ls5$ per cent. Based on these
results the IFs were combined in the image plane.

To investigate this thorny issue we define -- in overlapping regions
-- $\phi$ as the spherical angle formed by the source, the middle
distance between two pointings and the northerly direction. Flux
ratios are sensitive to pointing variations as a function of $\phi$,
as shown in Fig.~\ref{pri-beam1}. These diagrams can be used to prove
if the primary beam is really offset with respect to the phase centre.

In Fig.~\ref{pri-beam1} we plot $\phi$ as a function of the gain
factors $(I_{(1)}-I_{(2)})/(I_{(1)}+I_{(2)})$ and
$(S_{(1)}-S_{(2)})/(S_{(1)}+S_{(2)})$, where $I$ is the peak intensity
value corrected for bandwidth smearing and $S$ is the integrated flux
density (the subindexes $_{(1)}$ and $_{(2)}$ show the pointings being
compared). This estimate is based on single sources detected at ${\rm
PNR}>25$, and shows no substantial evidence for a primary beam
offset. In these plots, the signature expected for a primary beam
offset is a sinusoidal data distribution. To measure this effect we
use a sinusoidal $\chi^2$ fit, see dashed lines in
Fig.~\ref{pri-beam1}, to provide the {\em direction} of the offset via
the phase of the fit as well as the amplitude of the offset.

Based on this method to tackle the pointing offset, we find that the
use of peak flux values instead of integrated flux densities can
result in an {\it apparent} pointing offset, mostly because of the
$\chi^2$ fits obtained for the most widely separated pointings (3--1;
bottom in Fig.~\ref{pri-beam1}). Smearing effects may thus simulate
the behaviour expected for a primary beam offset when $I$ values are
used. Since we have accounted for bandwidth smearing, these fits (if
robust) suggest smearing is more prevalent than expected. This could
be due to inadequate bandpass calibration, but the cause is
fundamentally unknown. Since integrated flux densities ($S$) are not
affected by smearing effects, they should provide an unbiased
estimation for a primary beam offset. We show in the right-hand panels
of Fig.~\ref{pri-beam1}, that $\phi$ does not show compelling
signatures for primary beam offsets.

We ran simulations in order to test the reality of the small
amplitudes seen in the $\chi^2$ fits, applying different primary beam
corrections in the directions indicated by the fits, with the idea of
minimising the fitted amplitudes. We found that the offsets required
to remove the pointing offsets were $\ls\,1$\,arcmin, with different
directions for all the pointings. This contradicts the single offset
of $\sim\,$2.5\,arcmin, in a common direction, used by \citet{Garn07}.

Due to the lack of evidence for a significant and consistent pointing
shift, we decided not to apply any primary beam offset to our data.

\subsection{Multiple sources}
\label{mul_sou}

The definition and identification of multiple systems is a common
problem in radio astronomy, i.e.\ how many fitted Gaussian peaks in
the image are related to a single galaxy? This is especially difficult
for deep radio observations where extremely deep optical imaging is
required to identify the host galaxy. We can look at the image to find
obvious double-sided jets from bright, extended radio galaxies
(FR\,{\sc ii}), but this becomes more difficult at faint flux levels
for all angular scales, for obvious reasons.

The classification of multiple systems is also highly
resolution-dependent. A source adequately described by a single
Gaussian at 610\,MHz may require more than one component in the higher
resolution 1.4\,GHz data, confusing catalogues, number counts, and the
study of spectral indices. Later, we explicitly refer to 610-MHz- or
1.4-GHz-selected samples to avoid confusion.

\begin{figure}
\begin{center}
\includegraphics[scale=0.41]{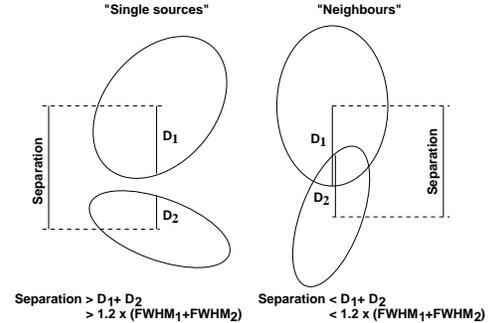}
\caption{A simple sketch for the selection criteria used to identity
  sources with close neighbours. If the source in the right were
  considered to be a double, then it would have an observed angular
  size given by `separation {\sc + (D$_1$ + D$_2$)/1.2}' (where D$_j$
  are 1.2 times the {\sc fwhm} in the direction to the neighbour
  source), and a position angle of 0\,deg.
  \label{sel_cri}
}
\end{center}
\end{figure}

In order to identify and classify the sources, we filtered the initial
3-$\sigma$ {\sc sad} catalogue, excluding all those fits with peak
values below $4\times$ the local noise. Then we identified all those
detections having close neighbours (with a ${\rm PNR}>4$) based on a
simple criterion: if the distance between two Gaussians is lower than
1.2$\times$ the sum of their measured {\sc fwhm}s in the direction
defined by the angle they form in the sky (see Fig.~\ref{sel_cri}),
then these detections are excluded from the so-called
``single-source'' sample. We have thus applied a
``friend-of-a-friend'' technique, using an elliptical
(direction-dependent) search radius -- an efficient classification
method. All these sources have been treated independently in order to
check their reliability. Only a small minority of them have been
considered as single emitters with more than one Gaussian
component. In Table~\ref{class} we describe the source
classifications. This identification is not restricted in flux density
\citep[e.g.][]{Ciliegi99, Seymour04, Biggs06} because some sources
display diffuse emission, or have fainter components superimposed. The
classification is presented in Tables~\ref{tableGMRT} and
\ref{tableVLA}, and is recommended to bear in mind when cross-matching
the catalogues.

\begin{table}
\scriptsize
\begin{center}
\begin{tabular}{|l|l|}
\hline
Class & Description \\
\hline
S  & Single Gaussian source without close neighbours  \\
SD & Single Gaussian source with one close neighbour \\
ST & Single Gaussian source with two close neighbours \\
SE & Single Gaussian source with multiple close neighbours \\
D  & Double Gaussian source without close neighbours \\
DT & Double Gaussian source with one close neighbour \\
T  & Triple Gaussian source without close neighbours \\
M  & Extended source composed of more than three Gaussians \\
\hline
\end{tabular}
\end{center}
\caption{Source classifications for the radio catalogues presented in
  Tables~\ref{tableGMRT} and \ref{tableVLA}. By `neighbours' we mean
  sources extracted by {\sc sad} with PNR\,$>4$, using the
  criteria shown in Fig.~\ref{sel_cri}.
\label{class}
}
\end{table}

{\sc tvstat} was used to determine the flux densities of extended
sources (usually sources with more than three Gaussian components --
see Table~\ref{class}). This gives a more accurate estimate for
complex systems than summing the various Gaussians. A final
inspection, by eye, checked the reliability of the components in
extended sources (including distant radio lobes in some cases),
sources showing side-lobe patterns and sources with diffuse emission
not included by the Gaussian fits. For all extended sources, errors in
peak and flux density are assumed to be 5 per cent of the value
reported by {\sc tvstat}.

Two other important source parameters are the observed angular size
and the orientation of the sources. We have measured the observed
angular size as follows. For multiple systems, it corresponds to the
separation of the furthest components, plus the measured {\sc fwhm} of
the components in the direction defined by the angle they form on the
sky (the `position angle'). For single sources, it is equivalent to
twice the maximum {\sc fwhm}. The observed angular size parameter and
the phase angle are presented in Tables~\ref{tableGMRT} and
\ref{tableVLA}.

\begin{table*}
\tiny
\begin{center}
\caption{A small sample of the sources found in the Lockman Hole field
at 610\,MHz using the GMRT. Source extraction is based on peak
brightness $>5\times$ the local {\rm r.m.s.} and integrated flux
density $>3\times$ the local {\rm r.m.s.} criteria. {\it Columns}: (1)
standard source name; (2) right ascension; (3) declination. We note in
\S\ref{astro_pres} there is an astrometric offset between the VLA with
respect to the GMRT sources, ${\Delta \rm R.A.= -0.60\,\pm\,0.03}$ and
${\Delta \rm Dec. = 0.40\,\pm\,0.03}$\,arcsec (mean offset in R.A.\
and Dec., respectively). For double and triple systems the position is
given by the brightest component. For extended sources it is given by
the most central component; (4) peak flux to local noise ratio; (5)
bandwidth smearing correction (\S\ref{subsub_bs}); (6) observed
maximum angular size (\S\ref{mul_sou}). These values are not
deconvolved source sizes but those fitted from the observed
mosaic. For single sources, this value corresponds to twice the
maximum {\sc fwhm}. For multiple sources, it is given by the distance
between the furthest components plus the {\sc fwhm}s of each of them,
in the direction they define in the sky; (7) The orientation angle
(position angle) of the source with respect to North; (8) integrated
flux density and estimated errors from {\sc sad}; (9) classification
of the source. S = single, D = double (d1 \& d2 as components), T =
triple (t1, t2 \& t3 as components) and M = extended
(Table~\ref{class}). The upper index $^{(\times2)}$ indicates sources
extracted from the convolved, Area$_{\rm beam}\times\sqrt{2}$, image
(\S\ref{scat}); (10) the radio spectral index between 610\,MHz and
1.4\,GHz, including the $68.3$ per cent confidence range (based on the
{\sc sad} flux density errors) and upper limits. `--' = outside
cross-matching region, `!' = unreliable spectral index; (11) Special
flags in spectral indexes: 1 -- upper limit, 2 -- source which has
split the counterpart's flux density in a relative contribution, 3 --
estimation which has used the original {\sc sad} extraction before
multiple classification, 4 -- cross-matched sources separated by a
distance $>3\,$arcsec, 5 -- counterparts having a fitted area ratio
twice bigger than the expected from point sources ($A_{\rm
source}^{\rm 610MHz}/A_{\rm source}^{\rm 1.4GHz}>2\times A_{\rm
beam}^{\rm 610MHz}/A_{\rm beam}^{\rm 1.4GHz}$), 6 -- estimation based
on a cross-match involving more than one counterpart, 7 -- spectral
index affected by close companion, and 8 -- source affected by
overlapping facets in the 3-D imaging approach. A complete version of
this table is available as Supplementary Material through the on-line
version of this paper.
\label{tableGMRT}}
\begin{tabular}{ccccccccccc}
\hline
IAU name & \multicolumn{2}{c}{Position at 610\,MHz (J2000)} & PNR  &
BWSC & Max.\ size & Position angle & Flux density & Class &
$\alpha^{\rm 610MHz}_{\rm 1.4GHz}$ & $\alpha$ flags\\ 
 & hr:min:sec & deg:min:sec &  &  & (arcsec) & (deg) &  ($\mu$Jy) &  &  &  \\
 (1) & (2) & (3) & (4) & (5) & (6) & (7) & (8) & (9) & (10) & (11)\\
\hline
GMRTLHJ105133.4+571459 & 10:51:33.42 & +57:14:59.9 &    7 & 0.99 &  14.3 &  56 &    105$\pm$   25 &  S & $-0.55^{+ 0.38}_{- 0.33}$ &  \\
GMRTLHJ105133.6+571308 & 10:51:33.59 & +57:13:08.8 &    9 & 0.99 &  19.1 &  97 &    188$\pm$   29 &  S & $-1.05^{+ 0.26}_{- 0.24}$ &  \\
GMRTLHJ105133.6+565039 & 10:51:33.64 & +56:50:39.1 &    5 & 0.96 &  17.8 &  89 &    185$\pm$   59 &  S & -- & --\\
GMRTLHJ105134.0+570552 & 10:51:33.96 & +57:05:52.4 &    8 & 0.98 &  14.6 &  83 &    116$\pm$   25 &  S & $-0.07^{+ 0.36}_{- 0.32}$ &  \\
GMRTLHJ105134.0+573729 & 10:51:34.02 & +57:37:29.2 &   11 & 0.97 &  15.8 &  71 &    242$\pm$   35 &  S & $-0.97^{+ 0.26}_{- 0.26}$ &  \\
GMRTLHJ105134.3+570922 & 10:51:34.31 & +57:09:22.6 &    7 & 0.99 &  13.9 &  65 &    177$\pm$   45 &  S & $ 0.81^{+ 0.37}_{- 0.29}$ &  \\
GMRTLHJ105134.5+573359 & 10:51:34.46 & +57:33:59.4 &   10 & 0.98 &  13.6 &  67 &    195$\pm$   34 &  S & $-0.52^{+ 0.29}_{- 0.26}$ &  \\
GMRTLHJ105134.5+573218 & 10:51:34.53 & +57:32:18.2 &    8 & 0.98 &  15.4 & 178 &    181$\pm$   36 &  S & $-1.46^{+ 0.37}_{- 0.36}$ &  \\
GMRTLHJ105134.6+574153 & 10:51:34.65 & +57:41:53.3 &    5 & 0.96 &  16.1 &   0 &    168$\pm$   46 &  S & $< -0.42$ & 1\\
GMRTLHJ105134.8+571801 & 10:51:34.81 & +57:18:01.6 &    7 & 0.99 &  17.5 &  65 &    134$\pm$   27 &  S & $-1.23^{+ 0.38}_{- 0.38}$ &  \\
GMRTLHJ105135.0+564615 & 10:51:35.02 & +56:46:15.7 &    7 & 0.94 &  16.1 &  51 &    325$\pm$   86 &  S & -- & --\\
GMRTLHJ105135.2+570133 & 10:51:35.19 & +57:01:33.9 &   10 & 0.98 &  17.4 &  15 &    234$\pm$   36 &  SE & $-0.80^{+ 0.33}_{- 0.36}$ &  \\
GMRTLHJ105135.3+570122 & 10:51:35.27 & +57:01:22.4 &    9 & 0.98 &  17.0 &  81 &    225$\pm$   36 &  SE & $< -0.71$ & 1\\
GMRTLHJ105135.6+572738 & 10:51:35.60 & +57:27:38.9 &   23 & 0.99 &  14.0 &  64 &    324$\pm$   24 &  S & $-0.69^{+ 0.12}_{- 0.11}$ &  \\
GMRTLHJ105135.6+570041 & 10:51:35.61 & +57:00:41.5 &    6 & 0.98 &  19.4 & 176 &    215$\pm$   49 &  S & $< -0.45$ & 1\\
GMRTLHJ105135.8+571344 & 10:51:35.84 & +57:13:45.0 &    6 & 0.99 &  22.6 &  47 &    145$\pm$   32 &  S & $-1.26^{+ 0.43}_{- 0.45}$ &  \\
GMRTLHJ105135.9+573728 & 10:51:35.92 & +57:37:28.1 &    6 & 0.97 &  13.5 &   7 &     92$\pm$   28 &  S & $-0.35^{+ 0.50}_{- 0.43}$ &  \\
GMRTLHJ105136.0+573424 & 10:51:36.05 & +57:34:24.7 &    5 & 0.98 &  13.3 &   7 &     73$\pm$   26 &  S & $< -0.31$ & 1\\
GMRTLHJ105136.1+574410 & 10:51:36.06 & +57:44:10.0 &    5 & 0.97 &  17.8 &  37 &    130$\pm$   38 &  S & $< -0.28$ & 1\\
GMRTLHJ105136.2+572959 & 10:51:36.20 & +57:29:59.1 &   27 & 0.99 &  15.3 &  72 &    605$\pm$   36 &  S & $-0.83^{+ 0.09}_{- 0.09}$ &  \\
GMRTLHJ105136.3+570651 & 10:51:36.28 & +57:06:51.2 &    9 & 0.98 &  13.2 &  41 &    132$\pm$   25 &  SD & $-0.96^{+ 0.39}_{- 0.42}$ &  \\
GMRTLHJ104809.1+570414 & 10:48:09.07 & +57:04:14.9 &    5 & 0.95 &  30.0 &  33 &    330$\pm$   90 &  S$^{\rm (\times2)}$ & -- & --\\
\hline
\end{tabular}
\end{center}
\end{table*}

\begin{table*}
\tiny
\caption{The 1.4-GHz catalogue. See Table~\ref{tableGMRT} for details.}
\vspace{0.2cm}
\begin{center}
\begin{tabular}{ccccccccccc}
\hline
IAU name & \multicolumn{2}{c}{Position at 1.4\,GHz (J2000)} & PNR  &
BWSC & Max.\ size & Position angle & Flux density & Class &
$\alpha^{\rm 610MHz}_{\rm 1.4GHz}$ & $\alpha$ flags\\ 
 & hr:min:sec & deg:min:sec &  &  & (arcsec) & (deg) &  ($\mu$Jy) &  &  &  \\
 (1) & (2) & (3) & (4) & (5) & (6) & (7) & (8) & (9) & (10) & (11)\\
\hline
VLALHJ105211.4+571551 & 10:52:11.44 & +57:15:51.7 &    5 & 0.95 &   8.3 & 178 &     28$\pm$    9 &  SD & $-0.53^{+ 0.40}_{- 0.50}\,!$ &   2\\
VLALHJ105211.5+573953 & 10:52:11.48 & +57:39:53.2 &   21 & 0.90 &   9.2 &  46 &    193$\pm$   17 &  S & $ 0.11^{+ 0.28}_{- 0.24}$ &  \\
VLALHJ105211.8+573510 & 10:52:11.82 & +57:35:10.2 &   12 & 0.94 &   7.1 &  37 &     53$\pm$    8 &  SD & $-1.55^{+ 0.35}_{- 0.32}\,!$ &   5\\
VLALHJ105211.9+570540 & 10:52:11.86 & +57:05:40.5 &    7 & 0.91 &   9.9 & 161 &     58$\pm$   16 &  S & $-0.22^{+ 0.50}_{- 0.49}$ &  \\
VLALHJ105212.0+572321 & 10:52:12.04 & +57:23:21.6 &    6 & 0.97 &   7.4 & 146 &     24$\pm$    7 &  S & $-1.48^{+ 0.54}_{- 0.54}$ &  \\
VLALHJ105212.1+573454 & 10:52:12.08 & +57:34:54.6 &    9 & 0.95 &   8.0 & 161 &     46$\pm$   10 &  S & $> -0.98$ & 1\\
VLALHJ105212.1+572621 & 10:52:12.11 & +57:26:21.4 &   10 & 0.97 &  10.2 & 136 &     64$\pm$   10 &  S & $-0.38^{+ 0.52}_{- 0.41}$ &  \\
VLALHJ105212.2+571525 & 10:52:12.16 & +57:15:25.4 &    9 & 0.94 &  10.4 &  73 &     68$\pm$   12 &  S & $-1.00^{+ 0.28}_{- 0.28}$ &  \\
VLALHJ105212.3+571549 & 10:52:12.27 & +57:15:49.5 &   19 & 0.95 &   8.0 &  90 &     84$\pm$    8 &  SD & $-0.53^{+ 0.21}_{- 0.20}\,!$ &   2\\
VLALHJ105212.5+572453 & 10:52:12.49 & +57:24:53.1 &   48 & 0.97 &   9.2 &  97 &    278$\pm$   10 &  S & $-0.62^{+ 0.10}_{- 0.09}$ &  \\
VLALHJ105212.6+570641 & 10:52:12.63 & +57:06:41.3 &    6 & 0.92 &  10.2 &  32 &     56$\pm$   17 &  S & $-1.61^{+ 0.41}_{- 0.48}$ &  \\
VLALHJ105213.3+572650 & 10:52:13.29 & +57:26:50.6 &   12 & 0.97 &   9.3 & 156 &     71$\pm$    9 &  S & $> -0.30$ & 1\\
VLALHJ105213.4+571605 & 10:52:13.38 & +57:16:05.3 &   52 & 0.95 &   8.8 &  59 &    301$\pm$   10 &  S & $-0.77^{+ 0.06}_{- 0.06}$ &  \\
VLALHJ105213.4+572600 & 10:52:13.44 & +57:26:00.2 &   10 & 0.97 &   7.8 & 161 &     45$\pm$    8 &  S & $-0.43^{+ 0.50}_{- 0.41}$ &  \\
VLALHJ105213.6+574436 & 10:52:13.64 & +57:44:36.0 &    6 & 0.89 &   7.8 & 166 &     64$\pm$   21 &  S & $> -0.93$ & 1\\
VLALHJ105213.8+571338 & 10:52:13.76 & +57:13:38.9 &   13 & 0.93 &   9.2 &   5 &     89$\pm$   11 &  SD & $-0.82^{+ 0.24}_{- 0.23}\,!$ &   2\\
VLALHJ105213.9+573935 & 10:52:13.89 & +57:39:35.9 &    9 & 0.90 &   7.4 & 124 &     54$\pm$   12 &  S & $> -0.95$ & 1\\
VLALHJ105214.0+571841 & 10:52:14.04 & +57:18:42.0 &    6 & 0.96 &  11.9 & 107 &     44$\pm$   11 &  S & $-1.32^{+ 0.42}_{- 0.44}$ &  \\
VLALHJ105214.2+573140 & 10:52:14.18 & +57:31:40.9 &    7 & 0.97 &  14.1 &  38 &     91$\pm$   17 &  S & $> -0.62$ & 1\\
VLALHJ105214.2+573328 & 10:52:14.21 & +57:33:28.2 &    7 & 0.96 &  11.9 &  35 &     85$\pm$   17 &  S & $-0.91^{+ 0.41}_{- 0.37}$ &  \\
VLALHJ105214.6+571335 & 10:52:14.60 & +57:13:35.9 &    5 & 0.93 &   7.9 &  62 &     24$\pm$    8 &  SD & $-0.82^{+ 0.42}_{- 0.54}\,!$ &   2\\
VLALHJ104914.4+570210 & 10:49:14.41 & +57:02:10.4 &    5 & 0.86 &  14.2 & 145 &    231$\pm$   69 &  S$^{\rm (\times2)}$ & $-0.56^{+ 0.36}_{- 0.44}$ &  \\
\hline
\end{tabular}
\end{center}
\label{tableVLA}
\end{table*}

\subsection{Source catalogues}
\label{scat}

Various clean-up processes were applied to the initial catalogues
produced by {\sc sad} (see \S\ref{sext}). First, we eliminated those
sources lying closer than 30 pixels (24 and 37.5\,arcsec at 1.4\,GHz
and 610\,MHz) from the image border, where the noise is considerably
higher, and removed sources forming part of multiple structures that
have been considered as single emitters. The final catalogues comprise
sources with PNR $>\,5$ (uncorrected by bandwidth smearing) and
integrated fluxes in excess of 3\,$\times$ the local {\rm r.m.s.}
(this avoids a small number of sources with extremely small sizes).

Lastly, we ran both source extractions again using
$\sqrt{2}\times$Area$_{\rm beam}$ convolved images to include extended
emission missed by the first selection process. We found 16 and 43 new
sources in the GMRT and VLA mosaics, respectively.

Final catalogues are presented in Tables~\ref{tableGMRT} and
\ref{tableVLA}. We have identified a total number of 1,587 and 1,452
sources with PNR $\ge$5$\sigma$ at 610\,MHz and 1.4\,GHz,
respectively. 

\subsection{Astrometric precision}
\label{astro_pres}

\begin{figure}
\begin{center}
\includegraphics[scale=0.34]{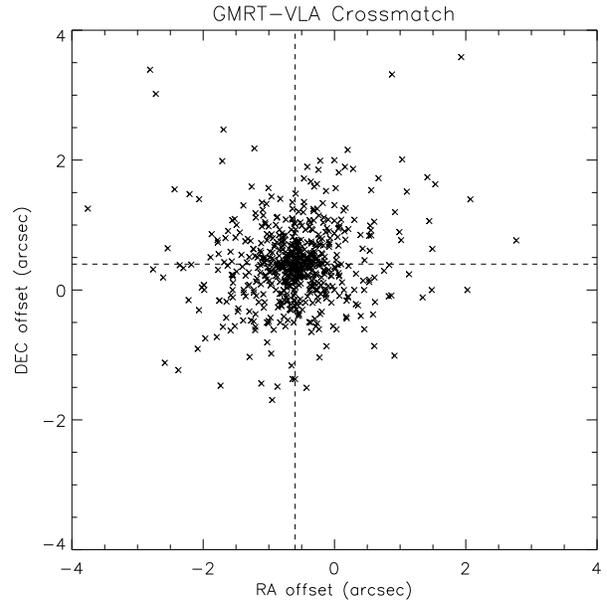}
\caption{Relative offsets between single sources found at 1.4\,GHz
  with respect to the 610-MHz GMRT positions. Offsets are
  approximately normally distributed. In R.A.\ and Dec.\ we find
  mean offsets of $-0.60\,\pm\,0.03$ and $0.40\,\pm\,0.03$\,arcsec,
  respectively. {\it Dashed lines} show the mean offset values.
\label{offsets}
}
\end{center}
\end{figure}

We plot in Fig.~\ref{offsets} the offsets of the VLA sources with
respect to the GMRT positions, using only single sources
(\S\ref{mul_sou}). The offset distribution is approximately Gaussian
with a mean of ${\Delta \rm R.A.= -0.60\,\pm\,0.03}$ and ${\Delta \rm
Dec. = 0.40\,\pm\,0.03}$\,arcsec.
These mean offsets, $\sim\,$0.5\,arcsec, are observed at all flux
levels, and their origin is unknown. \citet{Garn07} found an incorrect
time stamp in the GMRT data, resulting in a rotation of the positions
near the edge of each pointing.  This problem was corrected during the
reduction of our data and is not responsible for the observed offsets,
which may instead relate to VLA correlator issues which have only
recently come to light (\citealt{Morrison09}, in preparation), or be
due to a different position for the phase calibrators used at the
different frequencies.

An external reference was used to test the 610-MHz astrometry --
recent work by \citet{Garn08b}, which includes a 610-MHz observation
in the Lockman Hole. We find about 90 common sources (see
\S\ref{com_garn} for details) with a median offset of ${\Delta \rm
R.A.}=-0.54\pm0.05$ and ${\Delta\rm Dec.}=0.00\pm0.04$. These offsets
help explain the R.A.\ offset found in Fig.~\ref{offsets}, but not
that in Dec.

We have been unable to find a straightforward reason for the observed
offsets between the VLA 1.4-GHz and GMRT 610-MHz sources. We have not
implemented a positional shift in our catalogues, although we
highlight this issue in the table captions.

\section{Number counts}

We have derived number counts in the Lockman Hole using the catalogues
shown in Tables~\ref{tableGMRT} and \ref{tableVLA}. The differential
number counts, $dN/dS$, were calculated using the observed number of
sources per bin of flux density, $N$, divided by the bin width
($\Delta S$ in Jy) and by the effective area ($A_{\rm eff}$ in
steradians) available for detection.

\begin{equation}
\label{nc_equ}
  \frac{dN}{dS} = \frac{N}{A_{\rm eff}\Delta S}
\end{equation}

\subsection{Effective area}
\label{eff_are}

In our catalogues, the selection criteria for radio emitters is
determined by the local noise at the position of the source and by the
effectiveness of {\sc sad} in detecting the sources. In particular, we
note the noise across the image has a complicated structure and is
correlated on several different scales. In order to find the effective
area for source detection as a function of flux density, we modelled
25,000 point sources using the task {\sc immod}, with peaks from
1$\times$ to 500$\times$ the central r.m.s.\ in the mosaic
($\sim\,$15\,$\mu$Jy\,beam$^{-1}$ at 610\,MHz and
$\sim\,$6\,$\mu$Jy\,beam$^{-1}$ at 1.4\,GHz). We performed the
simulations by introducing 500 mock sources into a residual image
(without $\ge$5$\sigma$ positive or negative sources). We made use of
two different (but complementary) residual images obtained from the
negative (inverted) and the `positive' (normal) maps. This allowed us
to assess flux boosting due to random associations with faint, real
sources. We then extracted a catalogue using {\sc sad}, with the same
criteria as those described in \S\ref{cata}. Mock sources were
introduced, taking into account bandwidth smearing, with random
position angles, located $>$30\,pixels from the image border, none of
them overlapping. Bright residuals around powerful radio sources were
avoided too -- regions in the vicinity of the brightest 20
sources. The process of introducing sources was repeated 50 times.

Since mock sources were introduced randomly in the image, we can
assume that the completeness function -- the ratio of sources
extracted to those injected -- as a function of flux density,
normalised to the area of the field, provides the effective area,
$A_{\rm eff}(S)$, which includes all possible biases from the {\sc
sad} extraction and due to the noise structure of the map.

\begin{figure}
\begin{center}
\includegraphics[scale=0.34]{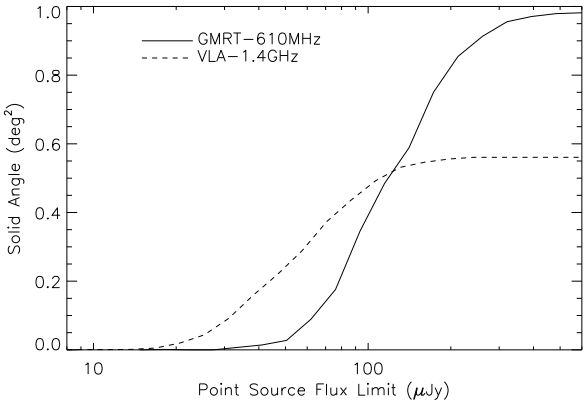}
\caption{Solid angle versus point-source flux limit.  {\it Solid} and
  {\it dashed} lines are based on GMRT and VLA mosaics, covering a
  total of 3,534 and 2,019\,arcmin$^2$, respectively.
\label{solid_angle}
}
\end{center}
\end{figure}

Fig.~\ref{solid_angle} shows the solid angle versus point-source flux
limit used in Eq.~(\ref{nc_equ}) to estimate the differential
number counts at 610\,MHz and 1.4\,GHz.

\subsection{Resolution bias}
\label{res-bias}

In this work, most of the sources lie in the sub-mJy regime
and the angular size distribution of the sub-mJy radio
population is not well known. Given the low resolution of our GMRT
and VLA maps, the vast majority of the sources are point-like
($\ls\,$4\,arcsec), therefore based on our data we are unable to
obtain a detailed and self-consistent angular size distribution.

Source catalogues based on a PNR-threshold criterion are biased
against extended sources as a function of flux density. In order to
estimate the fraction of sources being missed by our threshold
criterion, we applied the following treatment.

\subsubsection{Area threshold}
\label{ar_thres}

Number counts are calculated as a function of integrated flux density,
while our catalogues are mainly based on a PNR threshold
criterion. This restricts the detection of faint radio sources to
those with small angular sizes. The resolution bias can be simply
estimated using the following equation:

\begin{equation}
\label{flu_are}
  {\rm \frac{Flux_{[SAD]}}{Peak_{[SAD]}}} = 
  \frac{A_{\rm source}}{A_{\rm beam}} = 
    \left( \frac{\pi}{4\,\ln2} \right)\,\, 
    \frac{\rm Bmaj_{[SAD]}\,\,Bmin_{[SAD]}}{A_{\rm beam}}
\end{equation}

\noindent where the {\sc [sad]} index indicate values from one single
Gaussian from the {\sc sad} output fit; integrated `Flux' is given in
Jy, `Peak' flux in Jy\,beam$^{-1}$, Bmaj and Bmin are the fitted major
and minor {\sc fwhm}s in arcsec (not the deconvolved source size), and
the synthesised beam area, $A_{\rm beam} = \left( \frac{\pi}{4\,\ln2}
\right)\,{\rm Bmaj}^{\rm beam}\,{\rm Bmin}^{\rm beam}$\,arcsec$^2$.

Eq.~(\ref{flu_are}) implies that for our PNR threshold, the fitted
area of a source as a function of flux density is restricted to:

\begin{equation}
\label{thr_are}
  A_{\rm source} \leq \frac{A_{\rm beam}}{5\times {\rm r.m.s.}} {\rm
  Flux}
\end{equation}

In particular, our PNR threshold implies the faintest sources are
restricted to be detected with small fitted areas, $A_{\rm source}
\approx A_{\rm beam}$, or a little smaller than the beam given by the
uncertainties in the fit. Note that Eq.~(\ref{thr_are}) applies to
sources fitted with a single Gaussian only -- this is the case
for the vast majority of the sources analysed in this work.

\subsubsection{Angular size distribution}
\label{ang_sidis}

Previous studies have shown a decreasing angular size of the radio
sources towards faint flux densities. An early study,
\citet{Windhorst90} parameterised the angular size distribution of the
radio emitters using the following equation:

\begin{equation}
\label{res_are}
  h(\theta) = \exp[-\ln(2)\times(\theta/\theta_{\rm med})^{0.62}]
\end{equation}

\noindent where $h(\theta)$ is defined as the cumulative fraction of
sources with angular sizes larger than $\theta$ (the major {\sc fwhm})
and $\theta_{\rm med}=2\times\,S_{\rm 1.4GHz}^{0.3}$\,arcsec is the
median angular size as a function of flux density (in mJy). This
estimate predicts small variations in $\theta_{\rm med}$ as a function
of radio flux density. \citet{Bondi03}, however, found that the
\citeauthor{Windhorst90} distribution yields a considerably higher
number of sources with large angular sizes -- by almost a factor of
two (with $\theta>4$\,arcsec) than the observed in the sub-mJy
regime. This translates into an overestimate of the sources expected
to be missed in our observations. \citet{Bondi03} found that the
cumulative angular size distribution of sources with $0.4\leq S_{\rm
1.4GHz}<1\,$mJy (expected to be unbiased for $\theta\ls 15$-arcsec
sources) is well described by:

\begin{equation}
\label{res_are2}
  h(\theta) = 
  \begin{cases}
    (1.6^\theta)^{-1} & \quad {\rm for}\quad \theta\leq4\,{\rm arcsec}\\
    \theta^{-1.3}-0.01& \quad {\rm for}\quad \theta>4\,{\rm arcsec}.
  \end{cases}
\end{equation}

High-resolution radio observations (\citealt{Muxlow05, Biggs08}),
using data from both the Multi-Element Radio-Linked Interferometer
Network (MERLIN) and the VLA, have given us an angular size
distribution for the sources in the $40\,\mu{\rm Jy}<S_{\rm
1.4GHz}\ls\,1\,$mJy regime. Almost all the sources are resolved with
angular sizes below 4\,arcsec, which implies that our observations
might be unaffected by resolution bias at $\mu$Jy flux
densities. Nevertheless, \citeauthor{Muxlow05} estimate their sample
is 10 per cent incomplete based on previous observations with the
Westerbork Synthesis Radio Telescope (WSRT; with a synthesised
beamsize of 15\,arcsec, {\sc fwhm}) in the same field, which we use as
an upper limit for $\mu$Jy detections.

We have adopted a cumulative angular size distribution given by
Eq.~(\ref{res_are}) for bright ($>$1\,mJy) sources, and the average of
Eqs~(\ref{res_are}) and (\ref{res_are2}) for fainter sources.

\subsubsection{Correction factors}

In order to relate Eq.~(\ref{thr_are}) to the assumed cumulative
angular size distribution, we have considered $\theta={\sqrt{A_{\rm
source}/1.1331/\eta}}$, where $\eta$ is the median ratio between the
major and minor {\sc fwhm} (broadened by smearing effects) of the
observed VLA sources, $\eta=0.80$. Although this is a strong
assumption, changing this parameter does not significantly modify our
results.

\begin{figure}
\begin{center}
\includegraphics[scale=0.34]{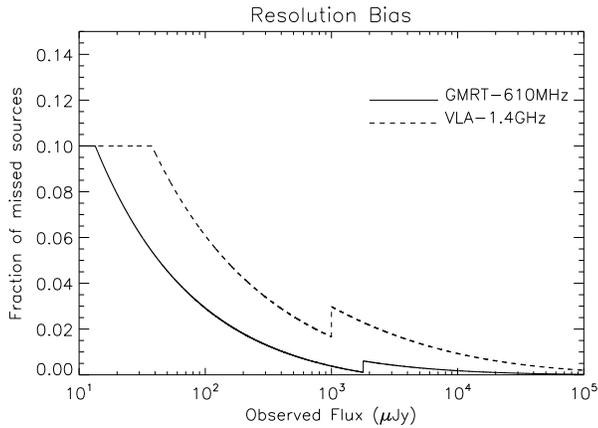}
\caption{Estimated missed fraction of extended sources as a function
  of flux density. {\it Solid} and {\it dashed} lines are the
  estimates for the VLA and GMRT detections, respectively. Estimates
  are based on Eq.~(\ref{thr_are}) and an averaged cumulative size
  distribution based on Eq.~(\ref{res_are}) and (\ref{res_are2}) --
  see \S\ref{ar_thres} and \S\ref{ang_sidis}. We have assumed a
  spectral index of $\alpha=-0.7$ for the GMRT predictions.
\label{resarea}
}
 \end{center}
\end{figure}

In Fig.~\ref{resarea} we plot the expected missed fractions as a
function of flux density for both observing frequencies. Note, we have
assumed a threshold of 10 per cent incompleteness 
for our faintest radio flux levels, based on \citet{Muxlow05}.

Since the angular size distribution of the 610-MHz sources is more
uncertain than that of the 1.4-GHz sources, we have assumed a radio
spectral index of $\alpha=-0.7$ to calculate the missed fraction at
610\,MHz.

These correction factors are small due to the relatively poor
resolution of our observations. Indeed, this bias is minimised when we
include extended sources extracted from the convolved images
(\S\ref{scat}).  In particular, at 100\,$\mu$Jy we predict that 3 per
cent of GMRT sources and 6 per cent of VLA sources are not selected in
our catalogues.

\subsection{Differential number counts}
\label{diff_nc}

\begin{figure}
\begin{center}
\includegraphics[scale=0.35]{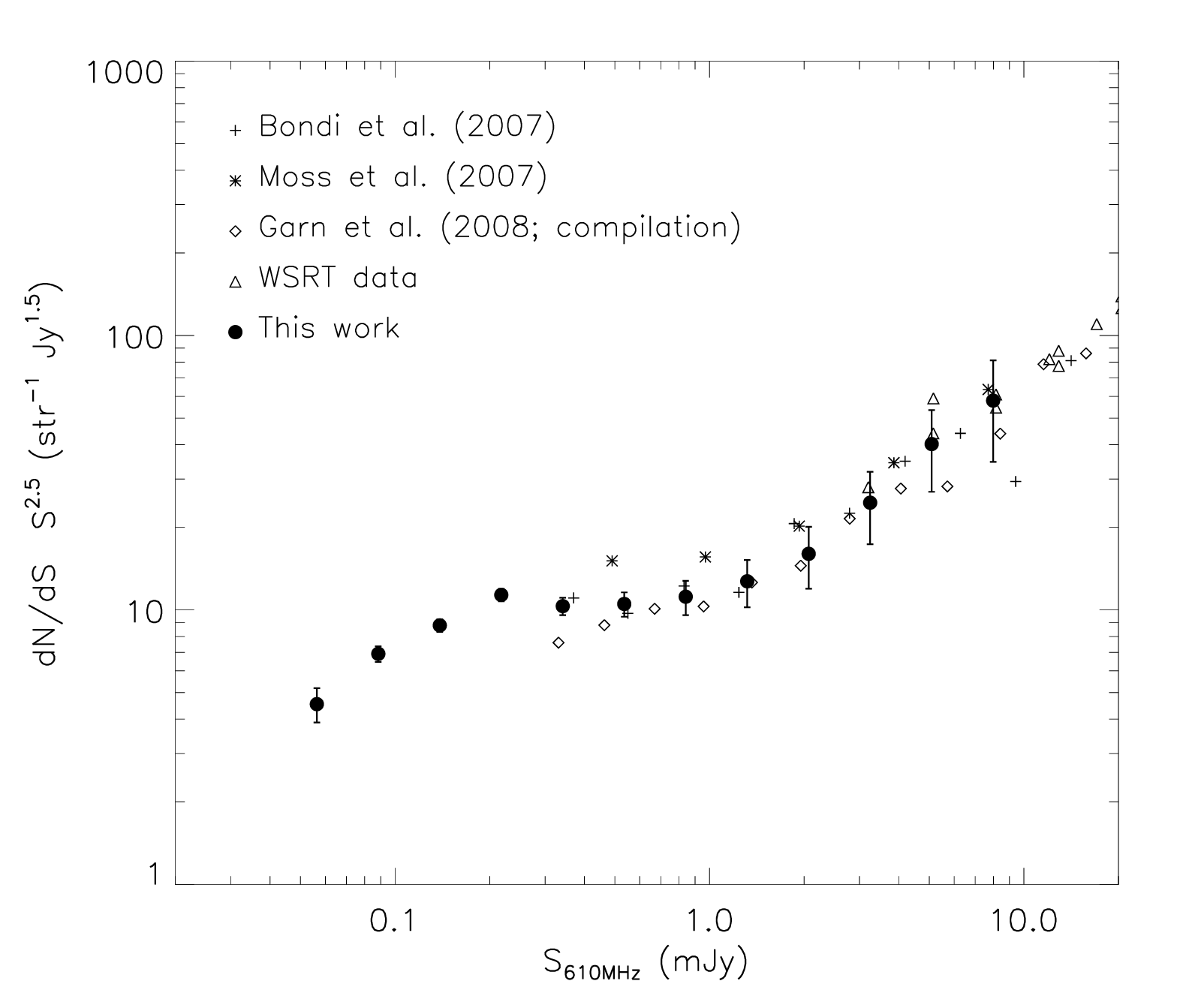}
\includegraphics[scale=0.35]{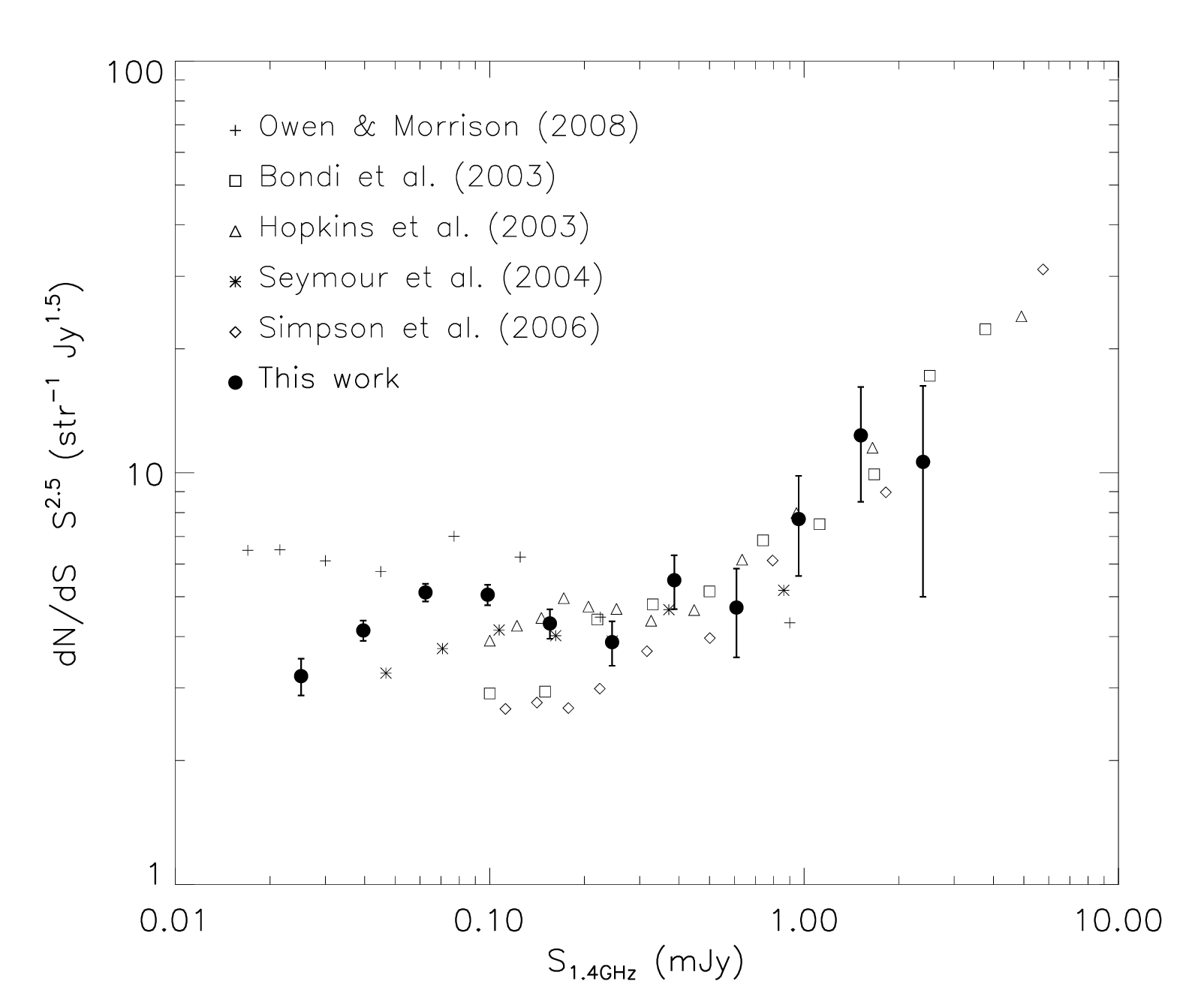}
\caption{Differential source counts as a function of flux density in
  the Lockman Hole at 610\,MHz ({\it top}) and 1.4\,GHz ({\it
  bottom}), normalised by the value expected in a static Euclidean
  Universe. Errors are assumed to be Poissonian \citep{Gehrels86} and
  are combined in quadrature for the observed number of sources in the
  bin and the mock source simulations described in
  \S\ref{eff_are}. At 610\,MHz, we plot data from previous studies:
  the VLA-VIRMOS Deep Field \citep{Bondi07}, the 1$^{\rm h}$ {\it
  XMM-Newton/Chandra} Survey \citep{Moss07}, a compilation from
  \citet{Garn08b} which includes data from the ELAIS N1, Lockman Hole
  and {\it Spitzer} First-Look Survey fields and counts obtained using
  the WSRT \citep{Valentijn77, Katgert79, Valentijn80, Katgert85}. We
  also show previous 1.4-GHz counts, based on studies using the VLA
  and using the Australia Telescope Compact Array: The Deep Swire
  Field \citep{Owen08}, the Phoenix Deep Survey \citep{Hopkins03},
  VLA-VIRMOS Deep Field \citep{Bondi03}, the 13$^{\rm h}${\it
  XMM-Newton/ROSAT} Deep X-ray Survey \citep{Seymour04} and the
  Subaru/{\em XMM-Newton} Deep Field \citep{Simpson06}.
\label{n_counts}
}
\end{center}
\end{figure}

The differential number counts from Eq.~(\ref{nc_equ}),
normalised for an Euclidean Universe, are plotted in
Fig.~\ref{n_counts}. At both frequencies we have used the effective
area shown in Fig.~\ref{solid_angle} and the correction for resolution
bias plotted in Fig.~\ref{resarea}.  The flux density used to
multiply the differential number counts is given by the bin centre (in
log space), and errors are Poissonian for uncorrected counts, for both
the observed and mock sources. Tables~\ref{tab_ncGMRT} and
\ref{tab_ncVLA} present the counts.

We observe a flattening in the Euclidean differential number counts
towards sub-mJy flux densities at both 610\,MHz and 1.4\,GHz
(Fig.~\ref{n_counts}). We find evidence for a second peak in number
counts at $\sim\,$80\,$\mu$Jy and $\sim\,$200\,$\mu$Jy for the VLA and
GMRT counts, respectively. These provide constraints on the
contribution of {\it IRAS}-like sources to the sub-mJy radio fluxes,
based on population synthesis models (\citealt{Hopkins00}). The
appearance of these features at sub-mJy radio fluxes is traditionally
explained as a transition from a dominant bright radio-loud AGN
population to a star-forming and radio-quiet AGN populations
\citep{Windhorst85, Simpson06, Condon07}.

In this work we extend the number counts down to very faint 610-MHz
flux densities whilst maintaining complete agreement with previous
studies at higher flux levels. Our results at 1.4\,GHz are in good
agreement with previous observations at $\gs200\,\mu$Jy using the
VLA's B configuration \citep{Bondi03, Seymour04} and the Australia
Telescope Compact Array \citep{Hopkins03}. At fainter levels,
$<100\,\mu$Jy, our 1.4-GHz counts are a little higher than the
majority of previous estimates. This may reflect underestimates of
number counts based on shallower images; for example, \citet{Owen08}
reported an approximately flat log\,$N$--log\,$S$ distribution down to
$S_{\rm 1.4GHz}\sim15\,\mu$Jy, exploiting an extremely deep VLA
1.4-GHz image of the 1046+59 field (r.m.s.\,$\sim$\,3\,$\mu$Jy). The
decrement of the number of radio sources towards faint flux densities
(Fig.~\ref{n_counts}) is highly dependent on the effective area,
$A_{\rm eff}$, in which detections are possible
(Fig.~\ref{solid_angle}). We find $A_{\rm eff}$ always decreases
slower in relation to the number of sources in each bin, so
underestimating our number counts would require us to have
undercorrected for resolution bias. We note, however, that even
adopting the \citet{Windhorst90} size distribution could not increase
the number counts to those found by \citeauthor{Owen08}.

A large number of faint radio sources was suggested recently by the
Absolute Radiometer for Cosmology, Astrophysics and Diffuse Emission
(ARCADE\,2 -- \citealt{Fixsen09}) experiment where an excess
brightness temperature was found in the 22\,MHz--10\,GHz range, where
the sky is expected to be dominated by synchrotron and free-free
emission from extra-galactic sources and the Milky
Way. \citet{Fixsen09} reported seeing approximately 5$\times$ the
expected contribution from faint radio sources (\citealt{Gervasi08})
to the cosmic microwave background, which sets a useful limit for the
total number of radio sources.

The origin of the wide scatter in reported 1.4-GHz number counts is
controversial. It is possible that for $<500\,\mu$Jy sources our
imaged area is not big enough to average out cosmic structure.  For a
field subtending an angle of one square degree, the angular diameter
distance at redshift unity is only $\sim$20\,Mpc. We note that
\citet{Condon07} estimated a count fluctuation of only
$\sigma=(1.07\pm0.26)\,N^{1/2}$ based on 17 non-overlapping fields in
the {\em Spitzer} First-Look Survey \citep[FLS --][]{Condon03}, where
$N^{1/2}$ is the statistical fluctuation expected without
clustering. Based on this, \citet{Condon07} stated that most of the
variance reported in the literature is `mundane, not cosmic', thereby
concluding that years of debate have been devoted to differences
induced by different instruments and analysis techniques (see
\S\ref{biggs_comp}) and possibly -- in more candid terms -- human
error.  \citet{Biggs06} came to a similar conclusion, finding around
double the source count in the {\em Hubble} Deep Field North as had
been measured by \citet{Richards00} and tracking the problem to a
simple arithmetical error rather than any fundamental problem with the
data or their reduction.

\begin{table}
\scriptsize
\caption{The 610-MHz radio source counts.}
\vspace{0.2cm}
\begin{center}
\begin{tabular}{ccccc}
\hline
$S$ bin & $S$ & $N$ & $ N/\Delta S/A_{\rm eff}$   & $dN/dS\times S ^{2.5} $\\
(mJy)   & (mJy)               &     & $({\rm str^{-1}Jy^{-1}})$   & $({\rm str^{-1}Jy^{1.5}})$\\
\hline
 0.045 -  0.071 &  0.056 &  58 & $(1.90\pm 0.27)\times10^{ 11}$&$  4.54\pm   0.65$\\
 0.071 -  0.111 &  0.088 & 253 & $(9.41\pm 0.61)\times10^{ 10}$&$  6.92\pm   0.45$\\
 0.111 -  0.174 &  0.139 & 379 & $(3.88\pm 0.20)\times10^{ 10}$&$  8.79\pm   0.46$\\
 0.174 -  0.273 &  0.218 & 389 & $(1.62\pm 0.09)\times10^{ 10}$&$ 11.35\pm   0.59$\\
 0.273 -  0.428 &  0.341 & 211 & $(4.79\pm 0.35)\times10^{ 9}$&$ 10.33\pm   0.75$\\
 0.428 -  0.671 &  0.536 & 114 & $(1.58\pm 0.16)\times10^{ 9}$&$ 10.52\pm   1.07$\\
 0.671 -  1.052 &  0.840 &  62 & $(5.46\pm 0.78)\times10^{ 8}$&$ 11.17\pm   1.60$\\
 1.052 -  1.651 &  1.318 &  36 & $(2.02\pm 0.39)\times10^{ 8}$&$ 12.72\pm   2.49$\\
 1.651 -  2.590 &  2.068 &  23 & $(8.24\pm 2.09)\times10^{ 7}$&$ 16.02\pm   4.07$\\
 2.590 -  4.063 &  3.244 &  18 & $(4.10\pm 1.21)\times10^{ 7}$&$ 24.60\pm   7.25$\\
 4.063 -  6.374 &  5.089 &  15 & $(2.18\pm 0.72)\times10^{ 7}$&$ 40.23\pm  13.29$\\
 6.374 - 10.000 &  7.984 &  11 & $(1.02\pm 0.41)\times10^{ 7}$&$ 57.92\pm  23.26$\\
\hline
\end{tabular}
\end{center}
\label{tab_ncGMRT}
\end{table}

Another possible origin for some of the reported scatter may be the
use of different VLA configurations for deep survey work, either
through problems setting the absolute flux scale, or via inadequate
correction for bandwidth smearing. We have separated the A- and
B-configuration data centred on our {\sc lockman-e} VLA pointing to
test if this introduces notable differences. We find a value of
$S_{\rm A}/S_{\rm B}=1.03\pm0.21$ for the mean and standard deviation
of the flux density ratios observed on sources detected in both
configurations with peak-to-noise ratios greater than 10 (42
sources). We conclude that use of different configurations does not
significantly bias estimates of flux densities in our catalogues. The
scatter must originate elsewhere, assuming that data have been
calibrated carefully, using appropriate $uv$ restrictions or
calibrator models.

Noise can contaminate the number counts through the inclusion of
spurious sources and by boosting intrinsically faint sources to higher
flux densities. We have tested the possibility that spurious sources
contaminate the samples by implementing an identical source extraction
procedure for the inverted (negative) signal map and we are confident
that any contamination by spurious sources is below 5 per cent, even
in the faintest flux density bins.

\begin{table}
\scriptsize
\caption{The 1.4-GHz radio source counts.}
\vspace{0.2cm}
\begin{center}
\begin{tabular}{ccccc}
\hline
$S$ bin & $ S $ & $N$ & $ N/\Delta S/A_{\rm eff}$   & $dN/dS\times S^{2.5} $\\
(mJy)   & (mJy)               &     & $({\rm str^{-1}Jy^{-1}})$   & $({\rm str^{-1}Jy^{1.5}})$\\
\hline
 0.020 -  0.032 &  0.025 &  95 & $(1.01\pm 0.10)\times10^{ 12}$&$  3.21\pm   0.33$\\
 0.032 -  0.050 &  0.040 & 284 & $(4.19\pm 0.24)\times10^{ 11}$&$  4.14\pm   0.24$\\
 0.050 -  0.078 &  0.062 & 388 & $(1.66\pm 0.08)\times10^{ 11}$&$  5.12\pm   0.25$\\
 0.078 -  0.124 &  0.098 & 303 & $(5.25\pm 0.30)\times10^{ 10}$&$  5.06\pm   0.29$\\
 0.124 -  0.195 &  0.155 & 158 & $(1.43\pm 0.12)\times10^{ 10}$&$  4.30\pm   0.35$\\
 0.195 -  0.308 &  0.245 &  76 & $(4.13\pm 0.51)\times10^{ 9}$&$  3.88\pm   0.48$\\
 0.308 -  0.485 &  0.386 &  55 & $(1.87\pm 0.28)\times10^{ 9}$&$  5.48\pm   0.82$\\
 0.485 -  0.765 &  0.609 &  24 & $(5.14\pm 1.25)\times10^{ 8}$&$  4.70\pm   1.14$\\
 0.765 -  1.206 &  0.961 &  20 & $(2.70\pm 0.74)\times10^{ 8}$&$  7.72\pm   2.11$\\
 1.206 -  1.902 &  1.515 &  16 & $(1.38\pm 0.43)\times10^{ 8}$&$ 12.33\pm   3.83$\\
 1.902 -  3.000 &  2.389 &   7 & $(3.81\pm 2.02)\times10^{ 7}$&$ 10.63\pm   5.63$\\
\hline
\end{tabular}
\end{center}
\label{tab_ncVLA}
\end{table}

\section{Spectral indices}
\label{spe_ind}

We have cross-matched (within 7\,arcsec) the GMRT and VLA radio
catalogues in order to estimate the spectral index of the radio
emitters. This measurement yields evidence for the synchrotron
mechanism which dominates the observed radiation from the sub-mJy
radio population. 

\begin{figure*}
\begin{center}
\includegraphics[scale=0.57]{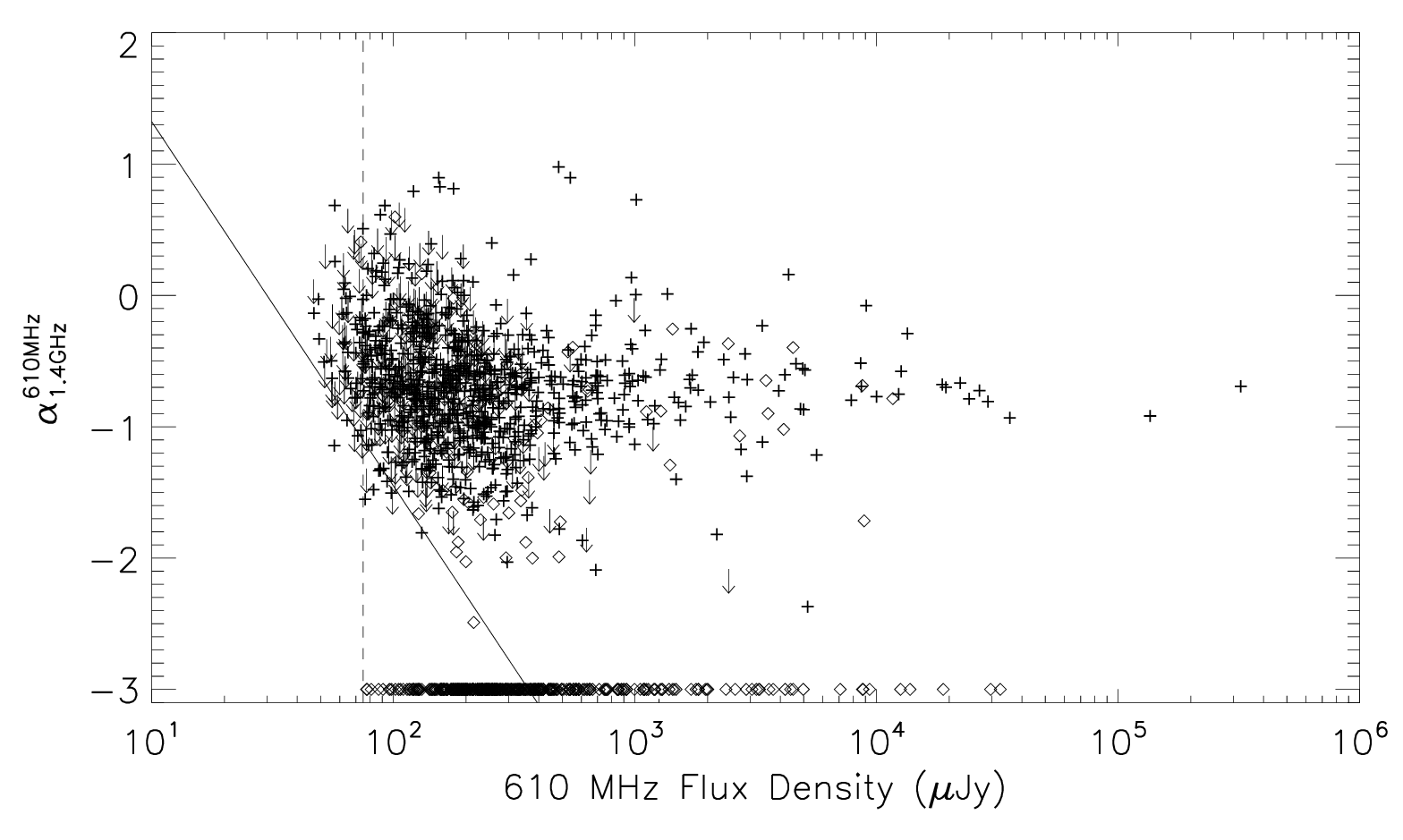}
\includegraphics[scale=0.57]{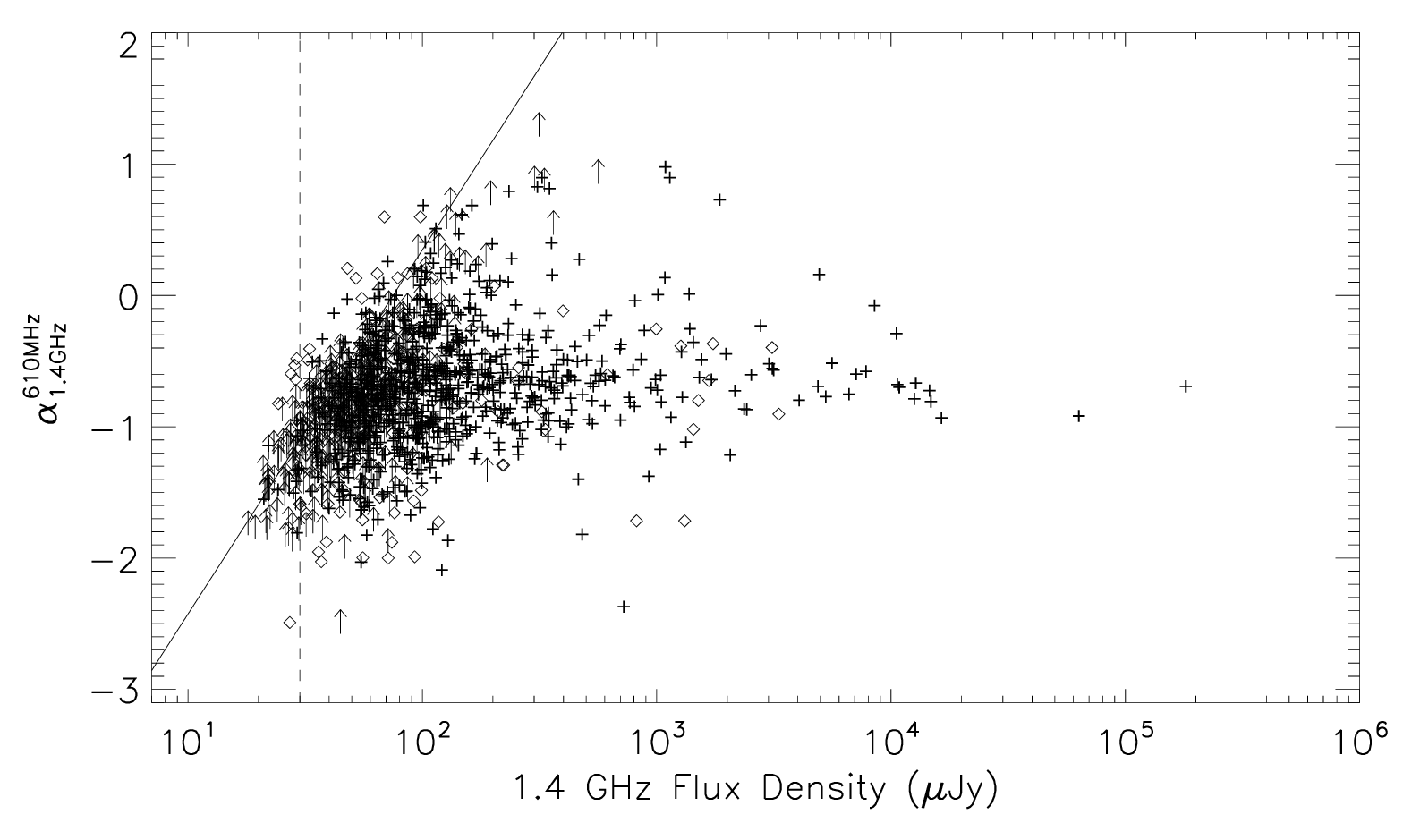}
\caption{Spectral index, $\alpha^{\rm 610MHz}_{\rm 1.4GHz}=-2.77\times
  {\rm log}(S_{\rm 610MHz}/S_{\rm 1.4GHz})$, as a function of flux
  density, based on a 610-MHz-selected sample ({\it top}) and a
  1.4-GHz-selected sample ({\it bottom}). {\it Dashed} and {\it solid}
  lines correspond to the primary and secondary point-source flux
  limit for each survey. {\it Plus} symbols represent detections. {\it
  Arrows} represent upper and lower limits for GMRT and VLA sources,
  respectively. {\it Diamonds} represent unreliable spectral indexes
  -- sources that were split, missing fainter components, or large
  offsets in the cross-match.  The data at $\alpha=-3$ are sources
  outside the overlapping region.
  \label{alpha}
}
\end{center}
\end{figure*}

In Fig.~\ref{alpha} we show the spectral index between 610\,MHz and
1.4\,GHz as a function of flux density.  We now analyse the spectral
indices based independently on GMRT- and VLA-selected samples, where a
610-MHz-selected catalogue naturally tends to prefer the detection of
steep-spectrum sources while selection at 1.4\,GHz favours flatter
spectra.

Since the VLA and GMRT images have different resolutions, special care
must be taken when we analyse results based on different
frequency-selected samples. The 610-MHz catalogue, from a
lower-resolution image, tends to have more counterparts per source
than the 1.4-GHz catalogue. This issue confuses the statistical
studies of spectral indexes since more unrelated galaxies are summed
up or split depending on the sample criterion.  For example, given our
selection criteria and the resolution difference between the observing
frequencies, it is possible that a single source at 610\,MHz is
related to two single detections at 1.4\,GHz.  In this case, the
spectral index of the GMRT source is calculated using the sum of the
flux densities from both VLA sources. For the VLA-selected sources we
have divided the flux density at 610\,MHz based on the relative
contribution from each VLA source (spectral indexes based on split
sources are flagged as such). Where there is no clear close
counterpart (within 7\,arcsec), 5-$\sigma$ upper/lower limits are
calculated using the local r.m.s.\ (\S\ref{sext}), weighting by the
source to beam area ratio.

Fig.~\ref{alpha} shows the spectral indices for GMRT-selected ({\it
top}) and VLA-selected ({\it bottom}) samples. Both distributions show
a large scatter, casting doubt on previous studies which assume a
clean star-forming galaxy population with a single spectral index in
the sub-mJy radio regime.

Upper/lower limits dominate at the faintest fluxes, partly due to the
difficulty in obtaining counterparts so close to the detection
threshold, where the catalogues are incomplete, and partly due to the
tendency to detect steeper-spectrum sources at 610\,MHz or
flatter-spectrum sources at 1.4\,GHz. In our work, this bias does not
allow the study of spectral index for sources with $S_{\rm 1.4GHz}\ls
100\,\mu$Jy. The larger number of lower limits at 1.4\,GHz reflects
the extra depth of the VLA imaging.

In Table~\ref{alp_res} we show the observed statistical results for
spectral indices as a function of flux density (parenthesis show mock
values -- see below). In order to avoid a large fraction of upper
limits in these statistical calculations, we have used only
$\ge$10-$\sigma$ (PNR) detections for each of the catalogue-based
samples, but down to $5$-$\sigma$ for the counterpart. We find no
trend in the distribution of spectral indexes toward fainter flux
densities in either the GMRT- or the VLA-selected catalogues. Since
these two samples tend to select spectra with different spectral
indexes, numerical differences of $\sim\,$0.2 in the mean $\alpha$,
and $\sim\,$0.1 in the median, are seen in the estimates for similar
flux density bins. These results contradict the suggested flattening
in spectral index at sub-mJy radio flux densities quoted in previous
studies \citep{Bondi07, Garn08a}.

\begin{table*}
\scriptsize
\centering
\begin{tabular}{|ccccc|}
\hline
\multicolumn{5}{c}{Based on the 610\,MHz-GMRT catalogue}\\
610\,MHz Flux Bin (mJy) & ${\rm \langle\alpha_{1.4GHz}^{610MHz}\rangle^{K-M}_{ASURV}}$ & ${\rm (\langle\alpha\rangle\pm\sigma_{\alpha})_{normal}}   $ & ${\rm (\langle\alpha\rangle\pm\sigma_{\alpha})_{biweight}} $ & ${\rm \langle\langle\alpha_{1.4GHz}^{610MHz}\rangle\rangle_{bootstrap}}$\\
\hline
$ 6.00-40.00$ & $-0.65\pm0.05$ & $-0.65\pm0.22\,\,(-0.70\pm0.01)$ & $-0.72\pm0.15\,\,(-0.70\pm0.01)$ & $-0.70\pm0.04\,\,(-0.70\pm0.00)$\\
$ 1.50- 6.00$ & $-0.83\pm0.09$ & $-0.78\pm0.47\,\,(-0.70\pm0.04)$ & $-0.70\pm0.33\,\,(-0.70\pm0.03)$ & $-0.68\pm0.06\,\,(-0.70\pm0.00)$\\
$ 0.50- 1.50$ & $-0.79\pm0.06$ & $-0.70\pm0.42\,\,(-0.74\pm0.23)$ & $-0.72\pm0.31\,\,(-0.71\pm0.20)$ & $-0.69\pm0.04\,\,(-0.72\pm0.00)$\\
$ 0.20- 0.50$ & $-0.91\pm0.03$ & $-0.85\pm0.40\,\,(-0.80\pm0.35)$ & $-0.85\pm0.38\,\,(-0.77\pm0.33)$ & $-0.85\pm0.04\,\,(-0.77\pm0.00)$\\
\hline
\multicolumn{5}{c}{Based on the 1.4\,GHz-VLA catalogue}\\
1.4\,GHz Flux Bin (mJy) & ${\rm \langle\alpha_{1.4GHz}^{610MHz}\rangle^{K-M}_{ASURV}}$ & ${\rm (\langle\alpha\rangle\pm\sigma_{\alpha})_{normal}}   $ & ${\rm (\langle\alpha\rangle\pm\sigma_{\alpha})_{biweight}} $ & ${\rm \langle\langle\alpha_{1.4GHz}^{610MHz}\rangle\rangle_{bootstrap}}$\\
\hline
$ 3.00-20.00$ & $-0.59\pm0.06$ & $-0.59\pm0.26\,\,(-0.68\pm0.02)$ & $-0.67\pm0.18\,\,(-0.68\pm0.02)$ & $-0.63\pm0.05\,\,(-0.68\pm0.00)$\\
$ 0.90- 3.00$ & $-0.48\pm0.11$ & $-0.48\pm0.59\,\,(-0.70\pm0.04)$ & $-0.64\pm0.50\,\,(-0.70\pm0.04)$ & $-0.60\pm0.09\,\,(-0.70\pm0.00)$\\
$ 0.30- 0.90$ & $-0.51\pm0.07$ & $-0.61\pm0.48\,\,(-0.68\pm0.18)$ & $-0.67\pm0.29\,\,(-0.69\pm0.17)$ & $-0.65\pm0.03\,\,(-0.69\pm0.00)$\\
$ 0.10- 0.30$ & $-0.48\pm0.03$ & $-0.57\pm0.43\,\,(-0.56\pm0.55)$ & $-0.59\pm0.41\,\,(-0.64\pm0.45)$ & $-0.60\pm0.02\,\,(-0.65\pm0.00)$\\
\hline
\end{tabular}
\caption{Statistical properties of radio spectral indices as a
  function of flux density. These estimates are based on confident
  measures of the spectral indices, i.e.\ with no flags in
  Table~\ref{tableGMRT} and \ref{tableVLA}. For both the 610-MHz- and
  1.4-GHz-selected samples we use only $\ge$10-$\sigma$ (PNR)
  detections (but down to 5-$\sigma$ for the counterpart) to minimise
  the uncertainties from upper/lower limits.  {\it Columns:} (1) the
  flux density bin; (2) the mean value given by the Kaplan-Meier
  product-limit estimator \citep[{\sc asurv} --][]{Feigelson85} which
  takes into account the upper limits in the bin; (3) the mean and
  standard deviation of the spectral index distribution based on
  detections only; (4) the central location (mean) and scale parameter
  (sigma) based on the biweight estimator which is resistant to
  outliers and non-Gaussian distributions (\citealt{Beers90}); (5) the
  median value of the distribution based on a bootstrap approach. The
  parentheses shown in some of the columns are the statistical
  estimates based on a single $\alpha=-0.7$ population with
  input/extracted flux ratios from the mock sources described in
  \S\ref{eff_are}. These constitute a useful check on the reliability
  of the results, e.g.\ whether the observed scatter is intrinsic or
  dominated by errors.
  \label{alp_res}}
\end{table*}

\begin{figure}
\begin{center}
\includegraphics[scale=0.35]{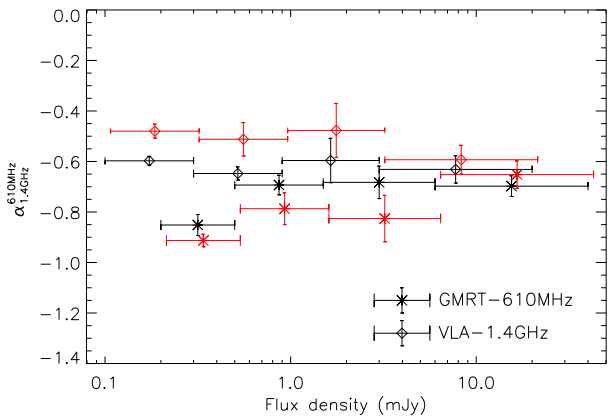}
\caption{The median-bootstrap ({\it in black}) and the Kaplan-Meier
  (\citealt{Feigelson85}) mean ({\it in red}) spectral indices as a
  function of flux density. Data are based on 610-MHz- and
  1.4-GHz-selected samples and shown in Table~\ref{alp_res}. In the
  image, mean values are slightly shifted in flux density (just for
  clarity).
  \label{meanalpha}}
\end{center}
\end{figure}

The almost constant mean and median values of $\alpha$ in the sub-mJy
regime (certainly between $0.1-10$\,mJy) is a robust result, and
suggests the sub-mJy radio population is dominated by optically-thin
synchrotron emission from star-forming galaxies and/or from
steep-spectrum lobe-dominated FR AGN. We show the trend for the mean
and median spectral index as a function of flux density in
Fig.~\ref{meanalpha}. The mean (in red) includes upper/lower limits
using the Kaplan-Meier product-limit estimator \citep[{\sc asurv}
--][]{Feigelson85}, and the median (in black) is obtained from
detections using a bootstrapping approach. The small variations in
these estimates reflect the non-Gaussianity and the large scatter of
the distribution, especially for the faintest detections. In
particular, the very steep spectral index obtained from the faintest
GMRT flux bin contains $\sim$20 per cent of upper limits, a result
which may be slightly biased toward steeper spectral indexes (see
simulations in parentheses from Table.~\ref{alp_res}).

The scatter of the $\alpha_{\rm 1.4GHz}^{\rm 610MHz}$ distribution is
$\sigma_{\alpha}\approx 0.4$ in the sub-mJy regime, which suggests the
detection of a large variety of populations -- probably a substantial
number of synchrotron self-absorbed AGN cores (\citealt{Blundell07,
Snellen00}) and high-redshift ultra-steep-spectrum (USS) sources
(\citealt{Jarvis01}).

Taking into account the point-source simulations from
\S\ref{eff_are}, we have estimated the distribution of spectral
indices based on variations between injected/extracted flux
densities. Assuming a radio population with $\alpha=-0.7$, and taking
into account the uncertainties in the source extraction process, we
predict that the spectral index distribution should broaden towards
faint fluxes, reaching a scatter similar to that observed in our
faintest flux density bins (see Table~\ref{alp_res}). These
simulations are presented in parenthesis next to the observed results,
and imply that the broad distribution of spectral indices is intrinsic
above $S_{\rm 1.4GHz}\gs\,100\,\mu$Jy but dominated (broadened) by the
fitting uncertainties (trumpet-like) at fainter flux densities.

Based on all sources, 6 per cent ($<$13 per cent) and 6 per cent ($<$9
per cent) have $\alpha>0$ in the GMRT- and VLA-selected catalogues,
respectively (the fractions in parenthesis include upper/lower
limits). Inspection by eye of these flat-spectrum sources reveals them
to be compact at both wavelengths, brighter in the VLA image than in
the GMRT image, probably due to synchrotron self-absorption in compact
($\ls$1\,kpc) GHz-peaked sources (\citealt{Snellen00}) which are
believed to be young FR\,{\sc ii} sources.

\section{X-ray identifications}

Deep, hard X-ray observations (in the 2--10\,keV band) provide an
efficient method for identifying AGN (\citealt{Mushotzky}), at least
with column densities $N_{\rm H}<10^{24}\,$cm$^{-2}$ (i.e.\ those that
are `Compton thin'). Heavily absorbed AGN are common
(\citealt{Maiolino}) and are expected to be responsible for the bulk
of the cosmic X-ray background (CXRB; \citealt{Ueda03, Hasinger04,
Gilli07}). Indeed, a large ($\sim$20--50 per cent) fraction is
believed to be Compton thick ($N_{\rm H}>10^{24}\,$cm$^{-2}$), a
population missed by even the deepest X-ray observations. Given that
radio observations are unaffected by obscuration, this suggests that
deep radio surveys may provide a method to find this missing
population -- the so-called QSO-2s.

\begin{figure*}
\begin{center}
    \includegraphics[scale=0.5]{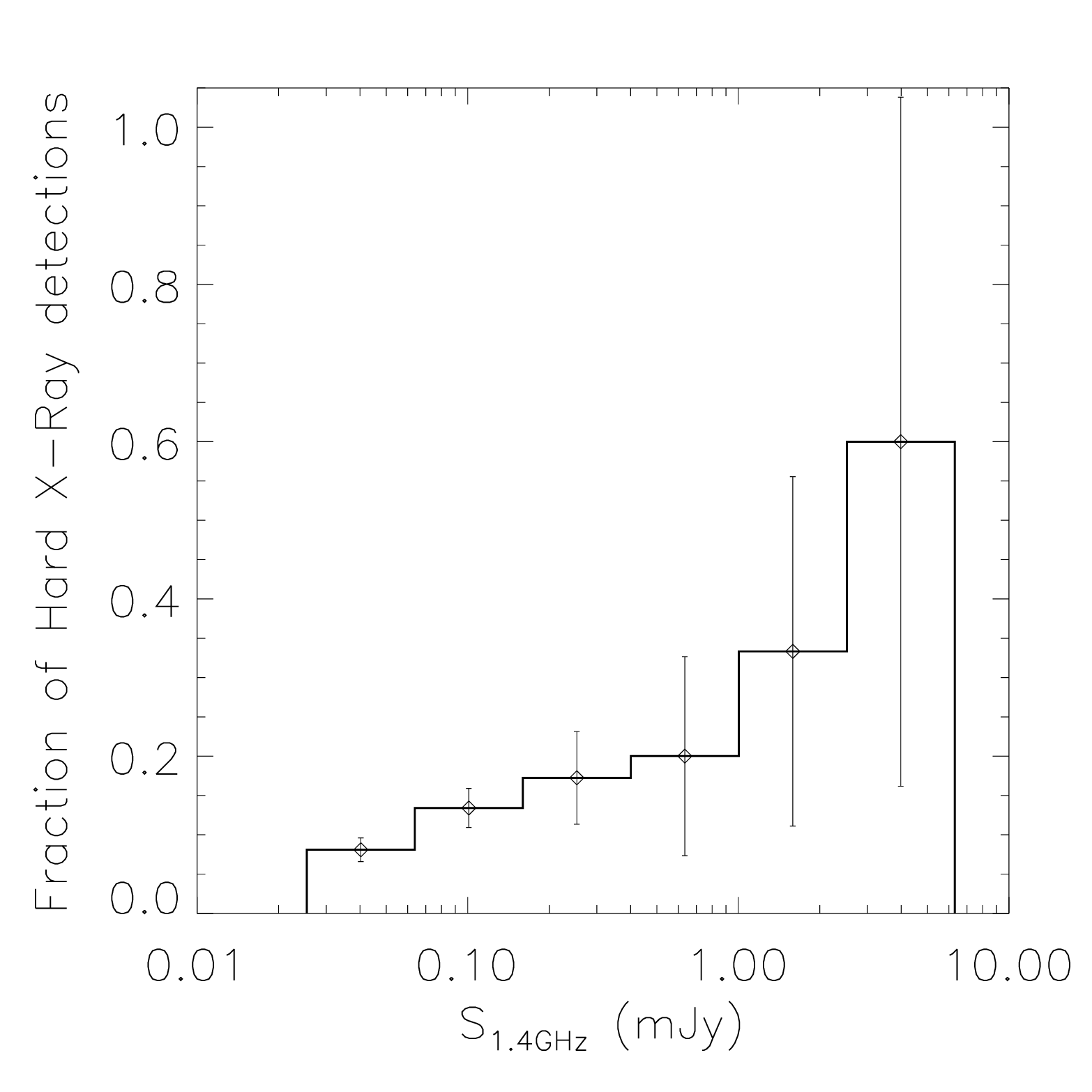}
    \includegraphics[scale=0.5]{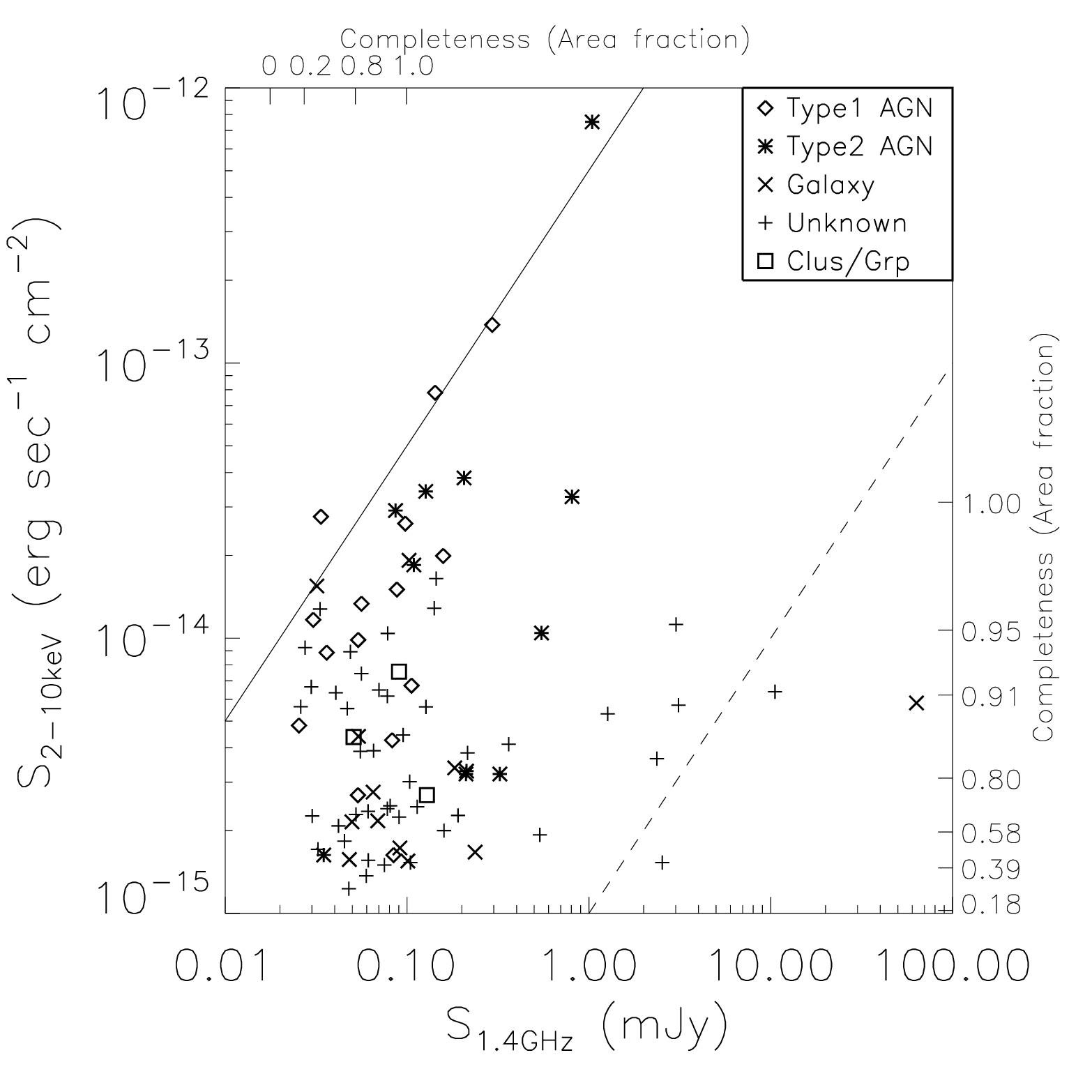}
    \caption{{\it Left:} The fraction of radio sources (at 1.4\,GHz)
    detected by {\it XMM/Newton} in the hard X-ray band. Errors are
    assumed to be Poissonian. {\it Right:} Radio flux density versus X-ray
    flux density for cross-matched sources. The solid and dashed lines
    correspond to the expected correlation for radio-quiet AGN
    (\citealt{Brinkmann00}) and star-forming galaxies
    (\citealt{Condon92,Ranalli03}) introduced by \citet{Simpson06},
    respectively. The different populations are based on spectroscopic
    identifications (\citealt[e.g.][in preparation]{Lehmann01,
    Szokoly09}) compiled by \citet{Brunner08}.
    \label{xray-fig1}
    }
\end{center}
\end{figure*}

In order to probe the nature of the sub-mJy radio population we have
cross-matched our VLA 1.4-GHz catalogue (Table~\ref{tableVLA}) with
the deepest {\it XMM/Newton} image so far published
\citep{Brunner08}. The X-ray field has an area of 0.196\,deg$^2$
(which is entirely covered by our VLA mosaic) and an effective
exposure time of 637\,ks. The X-ray catalogue contains 409 sources
above a likelihood of 10 (3.9\,$\sigma$) of which 266 and 340 are
detected in the hard and soft X-ray band ($S_{\rm
2-10keV}\geq9\times10^{-16} \,{\rm erg\, sec^{-1}\, cm^{-2}}$, $S_{\rm
0.5-2keV}\geq1.9\times10^{-16} \,{\rm erg\, sec^{-1}\, cm^{-2}}$),
respectively. The contamination by spurious sources in the X-ray
catalogue is expected to be only $\sim$1 per cent. 

In Fig.~\ref{xray-fig1} ({\em left}) we show the fraction of radio
sources detected in the hard X-ray band -- a good indicator for AGN
activity -- as a function of radio flux density. 32 per cent of the
hard X-ray sample (85 sources) have a $>5\sigma$ (PNR) 1.4-GHz
detection, within 5\,arcsec. This fraction declines from $\sim$30 per
cent at $\sim$1\,mJy to $\sim$10 per cent at $\ls100\,\mu$Jy. A large
number of faint radio sources, $S_{\rm 1.4GHz}<300\,\mu$Jy, that are
detected in the hard X-ray band. This has been previously noted by
\citet{Simpson06}, though in terms of the relative {\it fraction} they
remain a minority.

The deeper radio and X-ray catalogues used in this work -- compared to
those used by \citeauthor{Simpson06} (where $S_{\rm
1.4GHz}\geq100\,\mu$Jy and $S_{\rm 2-10keV}\gs3\times10^{-15}\, {\rm
erg\,sec^{-1}\,cm^{-2}}$ -- \citealt{Ueda08}) -- do not show any clear
evidence for a significant increase in the fraction of sources
detected in X-rays at faint radio fluxes. Indeed, when we bin the
X-ray data we find that between $S_{2-10\rm keV}=10^{-15} -
10^{-13}$\,erg\,sec$^{-1}$\,cm$^{-2}$, the fraction of X-ray sources
detected in the radio image is relatively constant, $\sim$35 per
cent. For the faintest X-ray sources, only 25 per cent are detected in
the radio image. We note that the use of a deeper, unpublished {\it
XMM/Newton} catalogue (Mat Page, {\it private communication}) does not
significantly modify the statistics at the faintest flux densities but
increases (to 50 per cent) the fraction of counterparts at
$S_{\rm 1.4GHz}\sim1\,$mJy.

The Lockman Hole is a popular legacy field and a large number of the
X-ray sources have been spectroscopically classified \citep[e.g.][in
preparation]{Lehmann01, Szokoly09}. We use the compilation of
\citet{Brunner08} to plot the hard X-ray and 1.4-GHz flux densities
(Fig.~\ref{xray-fig1}, {\em right}). Overplotted in the figure, we
show the expected correlations between radio and X-ray fluxes for
starbursts (\citealt{Condon92, Ranalli03}) and radio-quiet AGN
(\citealt{Brinkmann00}) as dashed and solid lines,
respectively. Deviations from these correlations can be produced by
photoelectric absorption (lower X-ray fluxes) or by jets oriented
close to our line of sight (larger radio fluxes).

We find that in the sub-mJy radio regime, the vast majority of the
radio-quiet AGN (types\,1 and 2) have $S_{2-10\rm keV}\gs
3\times10^{-15}$ \,erg \,sec$^{-1}$ \,cm$^{-2}$. This tallies with
\citeauthor{Simpson06}'s criterion for identifying radio-quiet AGN,
based on those sources lying closer to the solid than the dashed line
in Fig.~\ref{xray-fig1} ({\em right}). We find that the contamination
produced by galaxies at fainter X-ray flux densities could be very
large, since most of the spectroscopically identified galaxies lie at
$S_{2-10\rm keV}\ls4\times10^{-15}$ \,erg \,sec$^{-1}$ \,cm$^{-2}$ --
also in agreement with \citeauthor{Simpson06}'s criterion. Due to
photoelectric absorption, the distribution of type-2 AGN has a large
scatter in the hard X-ray band -- they are typically found with radio
fluxes, $S_{\rm 1.4GHz}\gs100\,\mu$Jy. It is interesting to note the
radio emission, at the $\sim$100-$\mu$Jy level, in two X-ray clusters
(XMMUJ105339.7+573520, XMMUJ105346.4+573510) and one classified group
(XMMUJ105318.9+572044).

From Fig.~\ref{xray-fig1} ({\em right}) we find that the solid line
(\citeauthor{Brinkmann00}'s correlation) appears to define an upper
limit for the AGN population. The well-known radio-loud AGN population
with $>1$\,mJy tends to agree better with the star-forming galaxy
correlation ({\it dashed line}). Finally, as the completeness
functions indicate -- {\it top} and {\it right} axes -- the faintest
radio/X-ray sources are observed in different areas of the map.

We have estimated the total fraction of radio-quiet AGN in the sub-mJy
radio regime based on four assumptions: (1) the X-ray catalogue
contains almost all the type-1 AGN in the redshift range of the radio
sources (mean, $z\approx 0.8$); (2) to estimate an upper
limit the spectroscopically-identified sources maintain the same
relative fractions in the unknown population as in
Fig.~\ref{xray-fig1} ({\it right}); (3) a constant fraction of
type\,1/type\,2 AGN\,$=$\,1:4 based on X-ray observations of local
Seyfert galaxies (e.g.\ \citealt{Maiolino}); (4) 25 per cent of the
X-ray sources are Compton-thick (undetected, obviously) based on
synthesis population models to reproduce the CXRB (e.g.\
\citealt{Ueda03}).

We find that the number of classified type\,1 AGN detected in both
1.4-GHz and X-ray wavebands is 21 (16 in the hard; 8 in the soft; 3 in
both). These sources comprise 38 and 31 per cent of the
spectroscopically classified samples (with radio detections) in the
hard and soft X-ray bands, respectively. Considering the assumptions
described above, we expect $\sim$131
$\left(21\times[1+4]\times[1+0.25] \right)$ AGN in our radio sample, i.e.
$\sim$20 per cent of the 755 radio sources in the region covered in
the X-ray waveband. Fig.~\ref{xray-fig1} ({\em right}) clearly shows
that most of the X-ray detections are at $S_{\rm
  1.4GHz}\ls300\,\mu$Jy, therefore applying the same treatment we
estimate that radio-quiet AGN compose a 19--37 and 19--30 per cent
fraction of the $S_{\rm 1.4GHz}<100\,\mu{\rm Jy}$ and $100\,\mu{\rm
  Jy}\leq S_{\rm 1.4GHz}<300\,\mu{\rm Jy}$ radio population,
respectively (upper limits based on the second assumption above). No
strong variations for the content of radio-quiet AGN as a function of
radio flux density are seen. These rough estimations are a little
higher than previous \citeauthor{Simpson06}'s estimate, i.e.\ 10--20
per cent.

These results suggest a transition, at sub-mJy radio flux levels, from
a bright and powerful AGN to a dominant star-forming galaxy
population, contaminated at the $\sim25\pm10$ per cent level by
radio-quiet AGN.

\begin{figure}
\begin{center}
    \includegraphics[scale=0.4]{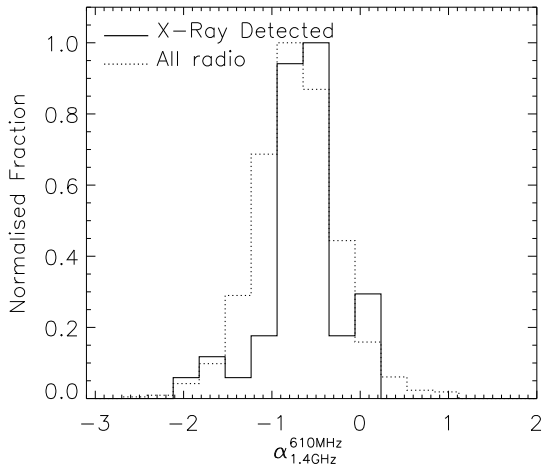}
    \caption{The radio spectral index between 610\,MHz and 1.4\,GHz
      for the sample of radio sources detected in the hard X-ray band.
    \label{xray-fig2}
    }
\end{center}
\end{figure}

Of the 84 hard X-ray sources detected at 1.4\,GHz, 48 have a reliable
radio spectral index (${\rm PNR}>5$ in both 610\,MHz and 1.4\,GHz
images). In Fig.~\ref{xray-fig2} we plot $\alpha^{\rm 610MHz}_{\rm
1.4GHz}$ for the X-ray sources alongside those of the entire VLA
sample. For the radio/X-ray sources, we find a median spectral index
of $-0.63\pm0.04$, slightly flatter than that of the whole 1.4-GHz
sample ($-0.70\pm0.01$). A Kolmogorov-Smirnov test gives a probability
of 16 per cent that both populations come from the same parent
distribution.  The spectral indices for the spectroscopically
identified X-ray populations are presented in
Table~\ref{spec_ident}. Given that most of the sources are found at
faint radio fluxes, these values may be highly biased by
incompleteness at 610\,MHz.

\begin{table}
\scriptsize
\begin{center}
\begin{tabular}{ccc}
\hline
Class & N & $\langle\langle\alpha^{\rm 610MHz}_{\rm 1.4GHz}\rangle\rangle$\\
\hline
Type\,1 AGN & 9 & $-0.79\pm0.20$ \\
Type\,2 AGN & 8 & $-0.60\pm0.12$ \\
Galaxy & 6 & $-0.65\pm0.26$ \\
Unknown & 24 & $-0.58\pm0.07$ \\
Cluster/Group & 1 & $-1.87$ \\
\hline
\end{tabular}
\end{center}
\caption{The median (bootstrap) radio spectral index for the
  spectroscopically identified hard X-ray sources plotted in
  Fig.~\ref{xray-fig1} ({\it right}).}
\label{spec_ident}
\end{table}

We conclude that radio-quiet AGN are no more numerous than
star-forming galaxies at faint flux densities. The fraction of radio
sources harbouring an AGN decreases towards faint radio flux densities
-- a transition from radio-loud AGN to a dominant star-forming galaxy
population at sub-mJy radio fluxes.

\section{Discussion}
\subsection{Comparison with a previous 610-MHz survey}
\label{com_garn}

\begin{figure}
\begin{center}
    \includegraphics[scale=0.34]{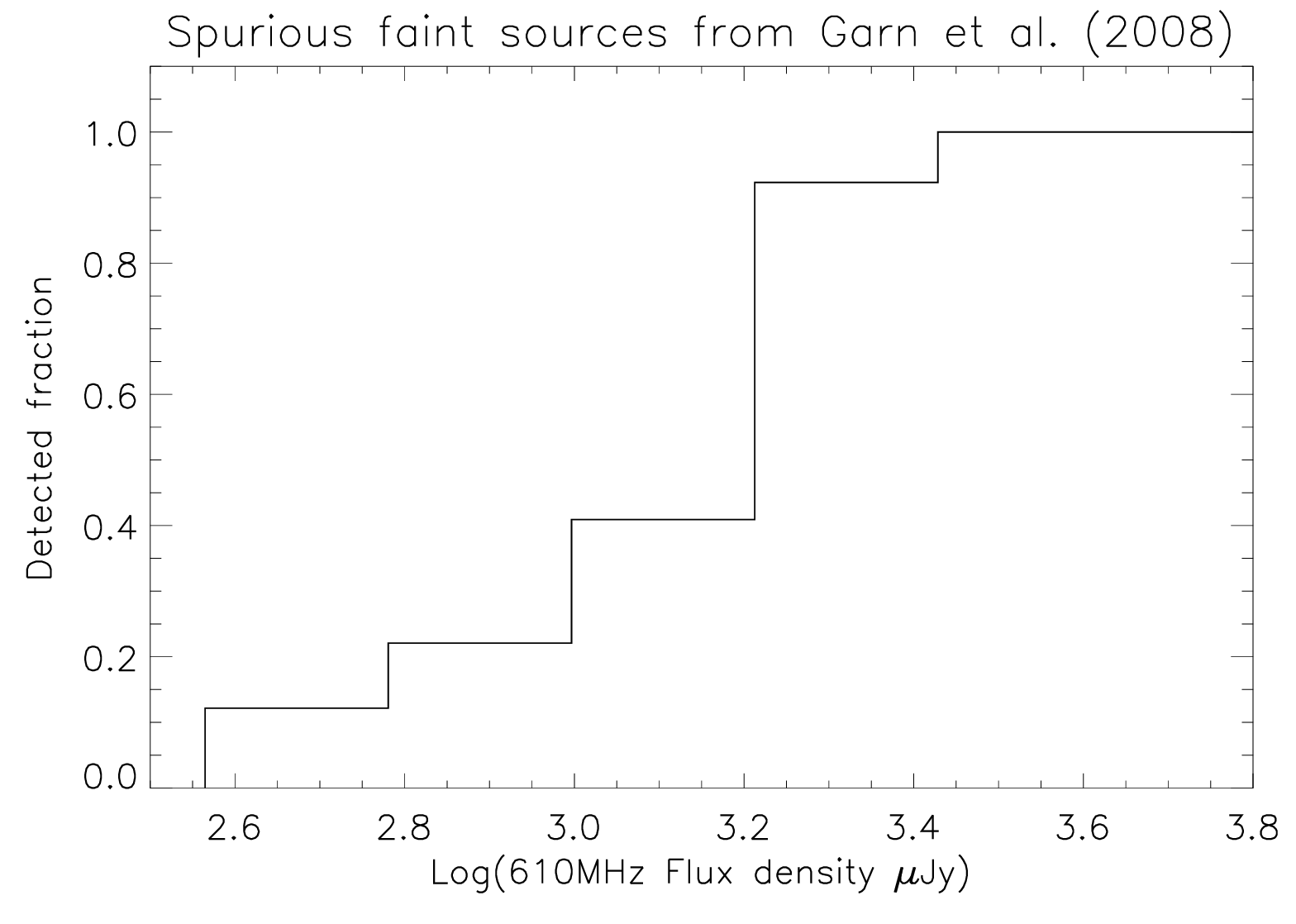}
    \includegraphics[scale=0.49]{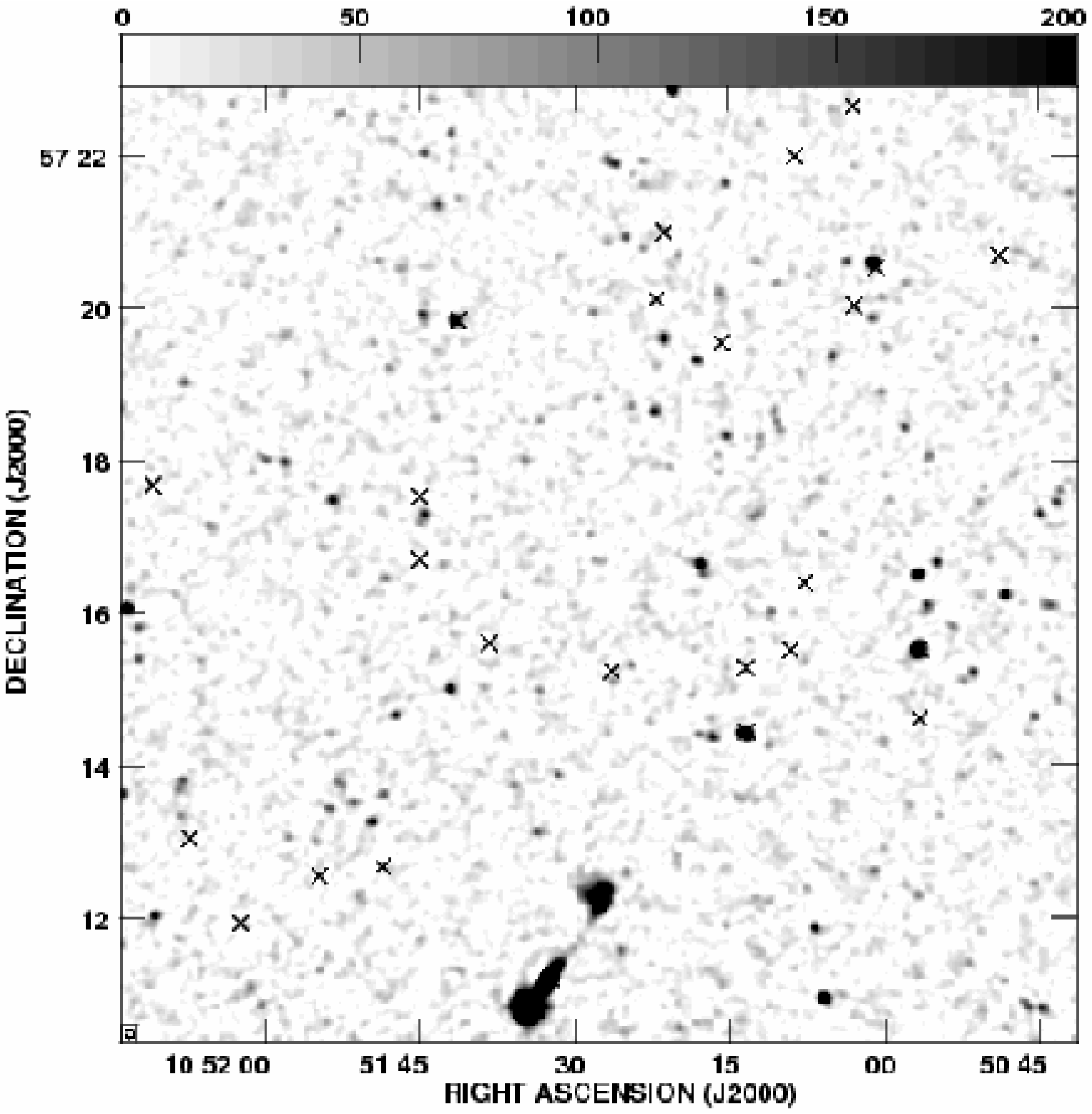}
    \caption{{\it Top:} Fraction of \citet{Garn08b} sources recovered
      in our image as a function of their 610-MHz flux density. {\it
      Bottom:} An area of $12\times12$\,arcmin$^2$ from our 610-MHz
      mosaic shown in grey-scale with a linear stretch from
      0--200\,$\mu$Jy\,beam$^{-1}$. Crosses show source positions
      catalogued by \citet{Garn08b}. We expect detections of all their
      sources since they have $S_{\rm 610MHz}^{\rm
      Garn}>300$\,$\mu$Jy.
    \label{fake-garn}
}
\end{center}
\end{figure}

\citet{Garn08b} published a 6-$\sigma$ 610-MHz catalogue covering
5\,deg$^2$ in the Lockman Hole using GMRT. This allows a direct
comparison with our detections at 610\,MHz. The \citeauthor{Garn08b}
mosaic ($\sigma_{\rm 610MHz}\sim\,$60\,$\mu$Jy\,beam$^{-1}$) covers
the north-western portion of our GMRT mosaic, with
$\sim\,$0.6\,deg$^2$ in common. Cross-matching their catalogue with
ours reveals good agreement in flux densities for the brightest
detections. However, we find a very large number of spurious, faint
sources in their catalogue. In Fig.~\ref{fake-garn} ({\em top}) we
show the fraction of \citeauthor{Garn08b} sources which are recovered
(within 7\,arcsec) in our 4$\times$ deeper image, as a function of
their radio flux densities. Given the depth of our image, we expect
all \citeauthor{Garn08b} sources should have a counterpart in our
catalogue, but no hints of emission in the image (Fig.~\ref{fake-garn}
-- bottom) are found for the vast majority of their $\ls\,3\,$mJy
(PNR\,$\ls\,10$) sources. We conclude that \citet{Garn08b} have not
provided a reliable 6-$\sigma$ catalogue. It is interesting to note,
however, that the number counts presented in Fig.~\ref{n_counts} are
in agreement with ours, probably because they base their number counts
considering bright sources ($\gs\,$10-$\sigma$) in the cleanest
regions of their map only. 

Looking at the full \citeauthor{Garn08b} Lockman Hole image, their
sources lie mainly near the edge of their pointings, where the noise
levels are greatest. The cleanest areas of their image contain very
few sources, suggesting that their source extraction did not utilise
knowledge of the local noise level. The distribution of 610-MHz
sources in the {\it Spitzer} FLS (\citealt{Garn07}, as used by
\citealt{Mag08}) appears similar, with a dearth of sources in the
deepest portion of the mosaic (see figure 1 of \citealt{Mag08}). This
suggests that some of the faint steep-spectrum emitters seen by
\citeauthor{Mag08} may be spurious (by random association), and that
their spectral index distribution is likely broadened at faint flux
densities.

\subsection{Comparison with a previous 1.4-GHz catalogue}
\label{biggs_comp}

\begin{figure}
  \begin{center}
    \includegraphics[scale=0.34]{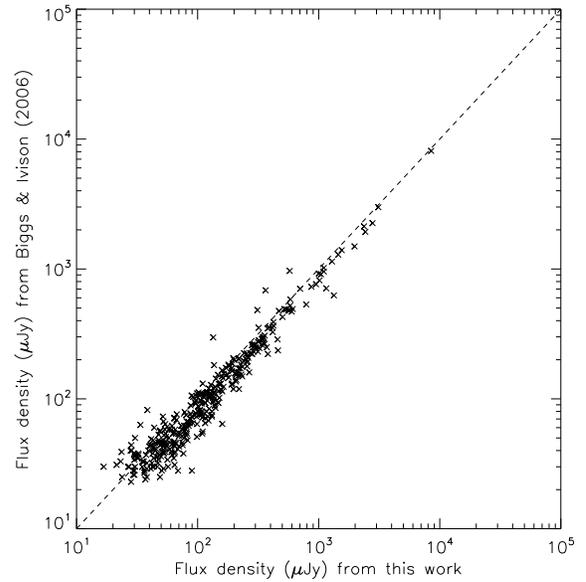}
    \caption{Integrated flux densities for the sources detected in the
      Lockman Hole at 1.4\,GHz in comparison with a previous work by
      \citet{Biggs06}. Sources have been cross-matched using a search
      radius of 4.2\,arcsec.
    \label{flux_biggs}
}
  \end{center}
\end{figure}

\citet{Biggs06} used some of the VLA data utilised in this paper. They
present a high-resolution (A-configuration) map with a
$\sim\,$1.3-arcsec beam ({\sc fwhm}), and a different method of source
detection. We find our work yields slightly higher flux densities.
The mean flux density ratio, $S_{\rm 1.4GHz}^{\rm ours}/S_{\rm
  1.4GHz}^{\rm Biggs}$, is $1.16\pm0.02$ (88 sources) with a standard
deviation of 0.15 for cross-matched sources with peak-to-noise ratio
higher than 15.  We have demonstrated in \S\ref{diff_nc} that this
difference is not produced by calibration problems from adding A- and
B-configuration data. A detailed analysis of both samples has been
carried out and we have found that differences in flux are
produced by the different approaches to source extraction. In particular,
\citet{Biggs06} used a fixed beam size to fit a Gaussian to sources
which were first extracted with areas smaller than the beam. This
results in lower measured flux densities, as seen in
Fig.~\ref{flux_biggs}. Reducing all our 1.4-GHz fluxes by a factor
1.16$\times$ would decrease the observed spectral indices
(numerically) by 0.18 (i.e.\ the spectra become steeper), but we
stress that this would not affect the (absence of) trend in the
spectral index distribution towards faint flux densities, nor the
width of the observed $\alpha$ distribution.

\subsection{Previous spectral index studies}

The radio spectral index observed between 610\,MHz and 1.4\,GHz has
been controversial since the earliest studies. Using the WSRT,
\citet{Katgert74} found -- from a small sample of sources with $S_{\rm
610MHz}\gs 10$\,mJy -- a spectral index distribution, $\alpha_{\rm
1.4GHz}^{\rm 610MHz}=-0.52\pm0.39$, an unusual result with a broad
distribution with respect to higher frequency surveys. A much larger
sample gave a similar result: $\alpha_{\rm 1.4GHz}^{\rm
610MHz}=-0.68\pm0.31$ \citep{Katgert79}, statistically in agreement
with the previous work, but showing clear evidence for more complex
than single power law spectra. Our survey covers only
$\sim\,$1\,deg$^2$ and therefore contains a small number (18) of
sources in the $S_{\rm 610MHz}\gs\, 10$\,mJy range. From these
sources, 12 have got a reliable radio spectral index from which we
find a mean $\langle\alpha\rangle\approx-0.71$ and standard deviation
of $\sigma_\alpha\approx0.17$, in agreement with early results but
with a considerably tighter distribution. Since these sources are
mostly powerful steep-spectrum radio-loud AGN (\citealt{Hopkins00}),
we have also compared these spectral indexes with a sample of $z<0.5$
FR {\sc ii} sources ($\alpha^{178MHz}_{750MHz}=0.79\pm0.14$, mean and
scatter), finding good agreements as well (\citealt{Laing83}).

With the advent of the GMRT, a variety of spectral index results have
appeared in the literature. They cover different flux density ranges,
so we present in Table~\ref{alpha_comp} a comparison using the same
ranges. We find good agreement -- within the errors -- with the
estimates of \citet{Bondi07} and \citet{Tasse07}, although not with
\citet{Garn08a}, probably because of faint spurious associations
(discussed in \S\ref{com_garn}) or incompleteness in their
estimations.

\begin{table}
\footnotesize
\begin{center}
\begin{tabular}{|ccc|}
\hline
Flux range (mJy) & 
$\langle\langle\alpha_{\rm 1.4GHz}^{\rm 610MHz}\rangle\rangle^{\rm Ref}$ &
$\langle\langle\alpha_{\rm 1.4GHz}^{\rm 610MHz}\rangle\rangle^{\rm Ours}$\\
\hline
$0.5<S_{\rm 1.4GHz}$      &$-0.67\pm0.05^{(1)}$ & 
$-0.63\pm0.03$ \\
$0.15<S_{\rm 1.4GHz}<0.5$ &$-0.46\pm0.03^{(1)}$ & 
$-0.63\pm0.02$ \\
$0.1<S_{\rm 1.4GHz}<0.15$ &$-0.61\pm0.04^{(1)}$ & 
$-0.59\pm0.03$ \\
\hline
$1.0<S_{\rm 610MHz}$     &$-0.56\pm0.04^{(2)}$ & 
$-0.69\pm0.04$ \\
$0.5<S_{\rm 610MHz}<1.0$ &$-0.36\pm0.12^{(2)}$ & 
$-0.71\pm0.04$ \\
\hline
$30<S_{\rm 610MHz}$ & $-0.76^{(3)}$ & 
$-0.71\pm0.17$ \\
\hline
\end{tabular}
\end{center}
\caption{A direct comparison of our results with previous work on the
  spectral index between 610\,MHz and 1.4\,GHz: (1) \citet{Bondi07};
  (2) \citet{Garn08a}; (3) \citet{Tasse07}. The double parentheses
  enclose median values, where we estimate our errors using a
  bootstrapping analysis.
\label{alpha_comp}}
\end{table}

\subsection{The nature of the sub-mJy radio emitters}

There is little agreement in the literature concerning the optical
properties of sub-mJy radio sources. Some studies support the idea
that faint radio emitters are primarily star-forming galaxies.  The
spectroscopic classification of $S_{\rm 1.4GHz}\gs\,40$-$\mu$Jy
sources by \citet{Barger07} shows that the dominant population has
strong Balmer lines and no broad or high-ionisation lines. Likewise,
\citet{Bondi07} find that late-type starbursts dominate the $S_{\rm
1.4GHz}\ls\, 100$\,$\mu$Jy regime based on analysis of an optical
colour-colour plot \citep{Ciliegi05}. Work based on a morphological
classification in the optical \citep{Padovani07} shows that
star-forming galaxies comprise only about a third of the sub-mJy
population. More radically, based on another optical colour-colour
study, \citet{Simpson06} presented evidence for no change in the
composition of the radio source population towards faint flux
densities, arguing for a dominant passively-evolving massive
elliptical galaxy population at all flux levels, $S_{\rm
1.4GHz}\geq100$\,$\mu$Jy. On the other hand, at $\sim\,$100\,$\mu$Jy,
bright submillimetre-selected galaxies -- which are very clearly
dominated by star formation -- make up a significant number of the
optically faint radio emitters \citep{Ivison02, Pope06}.

The approaches taken by all these studies reflect the difficulty of
disentangling star-forming galaxies from nuclear AGN activity.

We note that the large scatter ($\sigma_\alpha\approx0.4$) seen in
Fig.~\ref{alpha} for the spectral indices suggests a more complicated
scenario than a simple star-forming galaxy population (e.g.\ see
Fig.~\ref{xray-fig1}, {\em right}). We have found that the use of
radio spectral index as a probe of a galaxy's nature is highly
degenerate: supernova remnants and nuclear activity are closely
related (\citealt{Gebhardt00}); redshift effects may be combined with
synchrotron losses, steepening the spectra (\citealt{Jarvis01}) and --
given the poor resolution of our images -- spectral indices cannot be
obtained for resolved components. Therefore, a galaxy's nature is
difficult to disentangle using only the radio spectral
index. Nevertheless, in this work we find that the sub-mJy radio
population is characterised by optically-thin synchrotron emission,
contaminated at the $\sim25\pm10$ per cent level by radio-quiet AGN --
based on X-ray detections and previous spectroscopic
classifications -- in rough agreement with a previous study by
\citet{Simpson06}.

\section{Concluding remarks}

We have observed the Lockman Hole field using the GMRT at 610\,MHz and
the VLA at 1.4\,GHz, obtaining two deep radio images with similar
spatial resolutions and well-matched noise levels (15 and
6\,$\mu$Jy\,beam$^{-1}$, respectively) -- the former representing the
deepest GMRT image yet published. The data reveal a flattening
followed by a second peak in the Euclidean-normalised number counts in
the sub-mJy radio regime -- evidence for the appearance of a different
radio population dominating these faint flux densities.

We discuss the reliability of the 610-MHz catalogue presented by
\citet{Garn08b} in the Lockman Hole, finding that their catalogue is
highly contaminated by spurious sources, with similar problems
apparent in their previous {\it Spitzer} FLS catalogue
(\citealt{Garn07}). This may have influenced the detection of
steep-spectrum radio emitters reported by \citet{Mag08}.

We study the spectral index of the radio emitters by combining our
GMRT and the VLA measurements. The GMRT data are about $4\times$ times
deeper than previous imaging, allowing for clean results and avoiding
the well-known bias for steeper- or flatter-spectrum sources in
samples selected at longer or shorter wavelengths, respectively.
Analyses show evidence for large capabilities from GMRT observations
for imaging steep-spectrum steep-spectrum sources in the
field. Indeed, approximately 30 per cent of the GMRT sources are
undetected at 1.4\,GHz, revealing the sensitivity of deep 610-MHz
images to faint, high-redshift star-forming galaxies such as those
detected in submillimetre surveys \citep[e.g.][]{Ivison07}.

Our results, based on $\ge$10-$\sigma$ selection criterion, show that
the mean and median spectral index does not evolve as a function of
radio flux density, certainly between $\sim$100\,$\mu$Jy and 10\,mJy at
1.4\,GHz. We find $\alpha_{\rm 1.4GHz}^{\rm 610MHz}\approx -0.6$ to
$-0.7$, which suggests that optically-thin synchrotron emission is the
dominant emission mechanism in the sub-mJy population. The two most
probable contributors are star-forming galaxies and
\citeauthor{Fanaroff74} sources, ruling out a possible dominant
flat-spectrum population (AGN-cores; GPS) at these faint flux
densities (\citealt{Bondi07, Garn08a}).

We find the distribution of spectral indices has a
significant scatter ($\sigma_\alpha\approx0.4$), which suggests a
complicated scenario where different populations mix together. The
fraction of inverted-spectrum ($\alpha>0$) sources is just 6 per cent
($\ls$11 including lower limits) of the total sample.

Based on X-ray observations with spectroscopic classifications
(\citealt{Brunner08}), we estimate that approximately $\sim25\pm10$
per cent of the radio sample is made up of radio-quiet AGN at
$30\,\mu{\rm Jy} \ls S_{\rm 1.4GHz} < 300\,\mu$Jy, with this fraction
rising towards brighter flux densities. These results suggest a
transition from AGN to a dominant star-forming population at sub-mJy
radio fluxes.

\section*{Acknowledgements}

This paper was supported by a Gemini-STFC research studentship. We
thank the staff of the GMRT that made these observations possible.

\setlength{\labelwidth}{0pt} 
\bibliographystyle{mnras}
\bibliography{ibar}

\bsp

\end{document}